\documentclass[12pt,twocolumn]{emulateapj}

\usepackage{color}
\usepackage[usenames,dvipsnames]{xcolor}
\usepackage[colorlinks,urlcolor=blue,citecolor=blue,linkcolor=blue]{hyperref}
\usepackage[multidot]{grffile} 
\usepackage{amsmath}
\slugcomment{}

\begin{document} 

\title{Kinematics in the Galactic Bulge with APOGEE: II. High-Order Kinematical Moments and Comparison to Extragalactic Bar Diagnostics} 
\shorttitle{APOGEE Bulge Kinematics II}

\author{G.~Zasowski\altaffilmark{1}, 
M.~K.~Ness\altaffilmark{2},
A.~E.~Garc\'ia~P\'erez\altaffilmark{3,4}, \\
I.~Martinez-Valpuesta\altaffilmark{3,4},
J.~A.~Johnson\altaffilmark{5},
S.~R.~Majewski\altaffilmark{6}
}
\shortauthors{Zasowski et al.}

\altaffiltext{1}{Department of Physics \& Astronomy, Johns Hopkins University, Baltimore, MD 21218, USA; gail.zasowski@gmail.com}
\altaffiltext{2}{Max-Planck-Institut f\"ur Astronomie, D-69117 Heidelberg, Germany}
\altaffiltext{3}{Departamento de Astrof\'{i}sica, Universidad de La Laguna, E-38206 La Laguna, Tenerife, Spain}
\altaffiltext{4}{Instituto de Astrof\'{i}sica de Canarias, E-38206 La Laguna, Tenerife, Spain}
\altaffiltext{5}{Department of Astronomy, The Ohio State University, Columbus, OH 43210, USA}
\altaffiltext{6}{Department of Astronomy, University of Virginia, Charlottesville, VA, 22904, USA}

\begin{abstract}
Much of the inner Milky Way's (MW) global rotation and velocity dispersion patterns  
can be reproduced by models of secularly-evolved, bar-dominated bulges.
More sophisticated constraints, including the higher moments of the line-of-sight velocity distributions (LOSVDs)
and limits on the chemodynamical substructure,
are critical for interpreting observations of the unresolved inner regions of extragalactic systems and for placing the MW in context with other galaxies.
Here, we use SDSS-APOGEE data to develop these constraints, 
by presenting the first maps of the LOSVD skewness and kurtosis of metal-rich and metal-poor inner MW stars
(divided at ${\rm [Fe/H]} = -0.4$), 
and comparing the observed patterns to those that are seen both in $N$-body models and in extragalactic bars.  
Despite closely matching the mean velocity and dispersion, the models do not reproduce the observed LOSVD skewness patterns in different ways,
which demonstrates that our understanding of the detailed orbital structure of the inner MW remains an important regime for improvement.
We find evidence in the MW of the skewness-velocity correlation that is used as a diagnostic of extragalactic bar/bulges.
This correlation appears in metal-rich stars only, providing further evidence for different evolutionary histories of chemically differentiated populations.  
We connect these skewness measurements to previous work on high-velocity ``peaks'' in the inner Galaxy,
confirming the presence of that phenomenon,
and we quantify the cylindrical rotation of the inner Galaxy, 
finding that the latitude-independent rotation vanishes outside of $l \sim 7^\circ$.
Finally, we evaluate the MW data in light of select extragalactic bar diagnostics and
discuss progress and challenges of using the MW as a resolved analog of unresolved stellar populations.
\end{abstract}


\section{Introduction} \label{sec:intro}

Boxy/peanut bulges and bars dominate the inner regions of a large fraction of galaxies --- 
at least $\sim$45\% of massive disk galaxies \citep[e.g.,][]{Lutticke_2004_BPbulgestatistics,Buta_2015_S4Gmorphologies}.
These structures contain fossil records of the numerous events --- internal evolution and external interactions, both stochastic and secular events --- that have 
occurred throughout the lifetime of the galaxy \citep[see ][for a recent review]{Kormendy_2016_ellipticals-and-bulges}.
However, both dust extinction and perspective pose difficulties to interpreting these mostly-planar structures in edge-on systems.

The Milky Way (MW) presents a beautiful, close example of an edge-on boxy bulge that we can use 
not only to retrace the evolutionary history of our home galaxy, 
but also to characterize extragalactic edge-on bars and their impact on galactic properties.  
We can test the accuracy of extragalactic bar diagnostics using independent measurements of resolved stars, 
and understand the diversity of populations that can remain blended or hidden in those diagnostics of unresolved stars. 

Our understanding of the inner regions of the MW --- the boxy bulge or bar, and inner few kpc of the disk --- has been revolutionized
in the past several years by multiple surveys probing the large number of stars needed to measure the mean kinematical and chemical properties
of the highly extincted stellar populations \citep[e.g., BRAVA, ARGOS, GIBS, APOGEE;][respectively]{Rich_11_brava,Freeman_2013_argos,Zoccali_2014_GIBS1,Majewski_2015_apogeeoverview}.
The stellar population is dominated by metal-rich stars entrained in a thick bar with half-length $R_b \sim 2$~kpc, 
beyond which a thin planar bar component (with $R_b \sim 4.5$~kpc) is visible 
\citep[the ``long bar''; e.g.,][]{Zasowski_2012_innerMW,Benjamin_05_glimpse,Wegg_2015_RClongbar}.
The size, or even presence, of a ``classical'' bulge, resulting from the hierarchical merging of protogalaxies in the early days of the MW
or from in situ star formation during this same phase, remains a significant unknown
\citep[e.g.,][]{Shen_10_purediskbulge,Robin_2012_besanconbarmodel}.
See Section~4 of \citet{BlandHawthorn_2016_theMWproperties} for a recent review of our understanding of the inner MW.

\begin{figure*}[]
\begin{center}
  \includegraphics[trim=1.4in 1.7in 1.3in 4.0in, clip, angle=180, width=\textwidth]{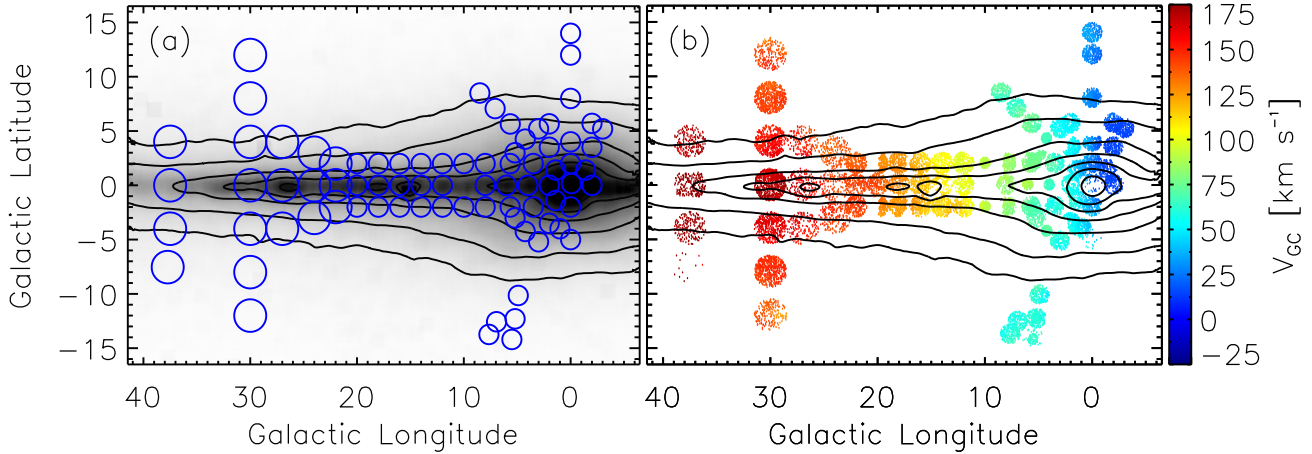} 
\end{center}
\caption{
Distribution of stars used in this analysis.  
(a) Positions and approximate sizes of the APOGEE fields (blue circles).
In grayscale is a smoothed [4.6$\mu$] WISE image of this region \citep{Lang_2014_unWISEcoadds}.  
Both the boxy shape of the bulge, and its asymmetry around the minor axis ($l=0^\circ$), are visible.
(b) Distribution of APOGEE stars against the WISE contours from panel (a), colored by the weighted mean velocities of neighboring stars.
}
\label{fig:WISE_datamap}
\end{figure*}

Empirically, the boxy bulge/bar displays roughly cylindrical  
rotation \citep[e.g., as seen by the BRAVA and ARGOS surveys;][]{Kunder_2012_bravaDR,Ness_2013_argoskinematics}.  
\citet[][hereafter Paper I]{Ness_2016_apogeekinematics} used APOGEE data to map this
mean rotation pattern and the velocity dispersion, which peaks in the center and falls off smoothly into the kinematically colder disk.
However, these properties are more correctly stated as describing the dominant metal-rich stellar populations;
the more metal poor stars show slightly different behavior 
\citep[slower rotation and higher velocity dispersion, known since, e.g.,][see also Paper I and \citet{Kunder_2016_bulgeRRLbrava}]{Harding_1993_bulgekinematics}.
These stars comprise a relatively small, but potentially very interesting, fraction of the total population 
\citep[e.g., $<$5\% have ${\rm [Fe/H]} \lesssim -1$;][A.~E.~Garc\'ia~Perez, in preparation]{Ness_2016_bulgeMDF}.

Numerous $N$-body models of the inner Galaxy have been developed and scaled to reproduce 
its mean rotation and velocity dispersion and
begin to explain the evolutionary history that has resulted in today's stellar chemodynamical patterns
\citep[e.g.,][]{Athanassoula_2007_barmodel,Shen_10_purediskbulge,MartinezValpuesta_2013_barmodel}.
Among the most important outstanding issues to address, through the combination of observations and these models,
are the mass fraction and dynamical impact of the long bar, the quantity and relevance of chemodynamical diversity in the inner Galaxy,
the mass contribution of the inner halo and any classical bulge,
and the relationships among kinematical, chemical, and morphological structures.
To address these questions --- to constrain and discriminate among models --- we need more than the mean rotation and dispersion patterns
that can be well-matched by models with a wide variety of initial conditions and histories.  
The {\it shape} of the kinematical distributions matters, along with the dependence of this shape on chemistry.

In this paper, we present the skew and kurtosis (the shape parameters) of the line-of-sight velocity distributions for $\sim$19,000 stars
observed predominantly in the inner $\sim$4~kpc of the MW.  
We discuss the implications for the claims of high velocity ``peaks'' in the inner Galaxy \citep{Nidever_2012_apogeebar} and 
compare these empirical results to those of MW-scaled $N$-body models and to common extragalactic bar diagnostics.  
We also examine the issue of cylindrical rotation in similar terms
as it has been quantified in external galaxies \citep[e.g.,][]{Molaeinezhad_2016_cylindricalBProtation}.
All of these are critical steps towards properly using the MW as a laboratory for understanding galaxy evolution at large.

\section{Data} \label{sec:data}

The Apache Point Observatory Galactic Evolution Experiment \citep[APOGEE;][]{Majewski_2015_apogeeoverview}
was a high resolution ($R=22,500$), near-infrared ($\lambda = 1.51-1.70$~$\mu$m) spectroscopic survey
to obtain radial velocities, fundamental parameters, and chemical abundances for over 100,000 red giant stars
sampling all components of the Milky Way \citep{Zasowski_2013_apogeetargeting,Holtzman_2015_apogeedata}.
Here, we use spectra and velocity data from Data Release 12 \citep[DR12;][]{Alam_2015_SDSSDR12}
of the Sloan Digital Sky Survey III \citep[SDSS-III;][]{Eisenstein_11_sdss3overview}.
APOGEE uses a 300-fiber spectrograph \citep{Wilson_2012_apogee} that, like all SDSS-III instruments, was  
coupled to the 2.5-meter Sloan Foundation telescope at Apache Point Observatory \citep{Gunn_2006_sloantelescope}.

Starting with the entire APOGEE sample contained in DR12, we first trim to an inner Galaxy sample with $|l| \le 45^\circ$ and $|b| \le 15^\circ$.
We remove stars only observed during survey commissioning, stars with unreliable RV measurements (${\rm VSCATTER} > 1$~km~s$^{-1}$ or
${\rm VERR\_MED} > 0.5$~km~s$^{-1}$), and stars with final ${\rm S/N} < 40$ per pixel.
We apply a $\log{g} \le 3.8$ requirement to remove the small number of foreground dwarf stars.  
Finally, we remove stars not selected as part of the ``normal'' sample --- e.g., stars observed with the NMSU 1-m telescope, ancillary targets,
telluric standards, cluster stars, calibration standards, and members of the Sgr dSph core or 
streams.\footnote{See \citet{Zasowski_2013_apogeetargeting} and \url{http://www.sdss.org/dr12/irspec/targets/} for APOGEE target information.}

\begin{figure*}[]
\begin{center}
  \includegraphics[trim=1.0in 1.0in 1in 1.1in, clip, angle=180, width=\textwidth]{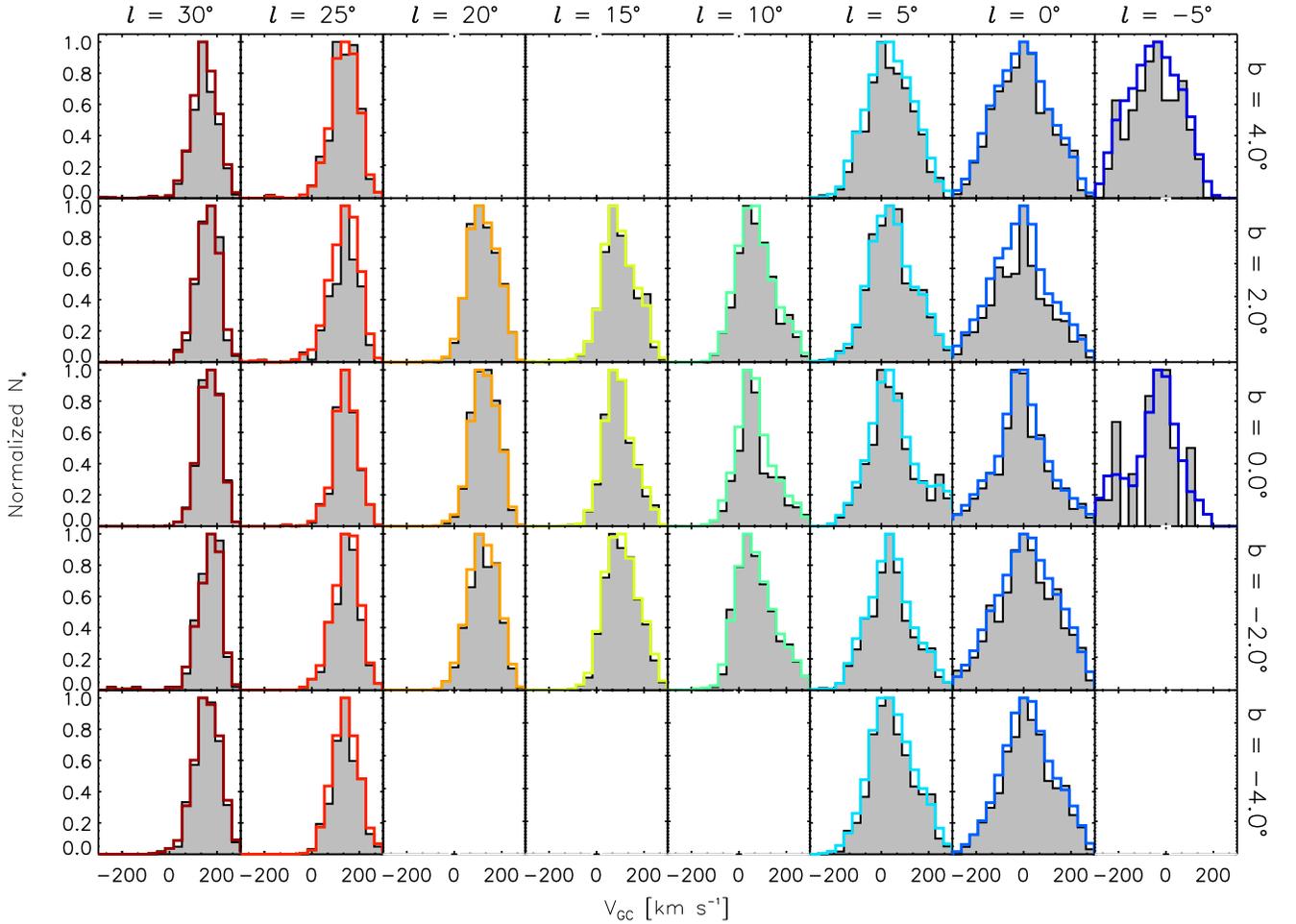} 
\end{center}
\caption{
Line-of-sight velocity distributions of the APOGEE stars used in this paper, 
arranged by Galactic longitude and latitude ($\Delta l = 5^\circ$, $\Delta b = 2^\circ$).
Gray histograms are the raw observed distributions, and the colored lines are the KDE distributions described in the text.
The number of stars in each populated bin ranges from 112 to 935, though we note that different binning schemes are used throughout the paper.
}
\label{fig:RV_histograms}
\end{figure*}

Radial velocities are calculated by APOGEE's data reduction pipeline \citep{Nidever_2015_apogeereduction}
and have a typical precision of $\sim$0.1~km~s$^{-1}$ for red giant stars.  
We use stellar parameters derived with {\it The Cannon} \citep{Ness_2015_Cannon},
computed from the DR12 spectra and trained with DR12 ASPCAP stellar parameters \citep{Holtzman_2015_apogeedata,GarciaPerez_2016_aspcap}.  
Distances are also derived from these {\it Cannon}-based parameters using isochrone-matching (as in Paper I),
with an estimated typical accuracy of $\sim$25\%.
We do not apply any global distance cuts, with the exception of requiring a reasonably converged solution
(which also serves to reject unphysical {\it Cannon} parameters), and we address the impact of this in the analysis (Section~\ref{sec:dist_cuts}).
With this limit, plus the other requirements above, our final sample contains 19,114 stars.

Figure~\ref{fig:WISE_datamap}a shows the sizes and distribution of the APOGEE field pointings against the [4.6$\mu$] flux density map
from WISE \citep{Wright_10_WISE,Lang_2014_unWISEcoadds}, which highlights the asymmetric boxy shape of the integrated starlight.
Figure~\ref{fig:WISE_datamap}b repeats the [4.6$\mu$] flux contours from \ref{fig:WISE_datamap}a, overplotted with the positions of the
stars used in this analysis.  
The color of each stellar point reflects the mean $V_{\rm GC}$ of its neighbors, weighted with a $\sigma = 0.5^\circ$ Gaussian kernel.

For all of the radial velocities used here, we transform between 
$V_{\rm helio}$ and Galactocentric  $V_{\rm GC}$ using:
\begin{equation}
\begin{aligned}
V_{\rm GC} = V_{\rm helio} + & 220 \sin{(l)} \cos{(b)} \\
+ & 16.5( \sin{(b)} \sin{(25^\circ)} \\
+ & \cos{(b)}\cos{(25^\circ)}\cos{(l-53^\circ)})
\end{aligned}
\end{equation}
adopting 220~km~s$^{-1}$ as the local standard of rest velocity and 
a solar peculiar velocity of 16.5~km~s$^{-1}$ towards $(l,b) = (53^\circ,25^\circ)$ \citep{MihalasBinney_1981},
though the analysis described here does not depend on the exact choice of these values.

\section{Velocity Moments} \label{sec:raw_VDs}

We directly calculate the first four moments of the velocity distributions using the standard definition of the 
moments\footnote{That is, for a velocity distribution of $N(V)$: 
mean $<$$V$$>$ or $\mu = \int{VN(V)dV}$, dispersion $\sigma(V)=\sqrt{\int{(V-\mu)^2N(V)dV}}$, 
skewness ${\rm Skew}(V) = \int{(\frac{V-\mu}{\sigma})^3 N(V)dV}$, and kurtosis ${\rm Kurt}(V) = \int{(\frac{V-\mu}{\sigma})^4 N(V)dV}-3$.}
(i.e., not fitting a Gauss-Hermite series or other parameterization, as is common in the extragalactic literature).
To ameliorate biases due to particular choices of histogram bin widths and locations,
we perform these calculations on a kernel density estimator (KDE) of the velocity distribution,
with a Gaussian kernel function and a bandwidth $h$ that is tuned to each distribution 
with the prescription $h=1.06\sigma N_*^{-0.2}$ 
\citep[][where $\sigma$ is the dispersion and $N_*$ is the total number of stars]{Silverman_1986_densityestimation}.
Visual inspection of the velocity distributions with a variety of histogram bin choices
confirms that the distributions are normal enough for this prescription to be applicable.
Figure~\ref{fig:RV_histograms} shows the velocity histograms (with one arbitrary choice of bin spacing)
and the KDEs for stars in several ranges of longitude and latitude.
The variations in velocity distribution centers, widths, and asymmetries are explored in detail 
in Sections~\ref{sec:low_moments}--\ref{sec:high_moments}.

\subsection{Velocity Mean and Dispersion} \label{sec:low_moments}

\begin{figure*}[]
\begin{center}
  \includegraphics[trim=1in 0.9in 0.5in 0.7in, clip, angle=180, width=0.45\textwidth]{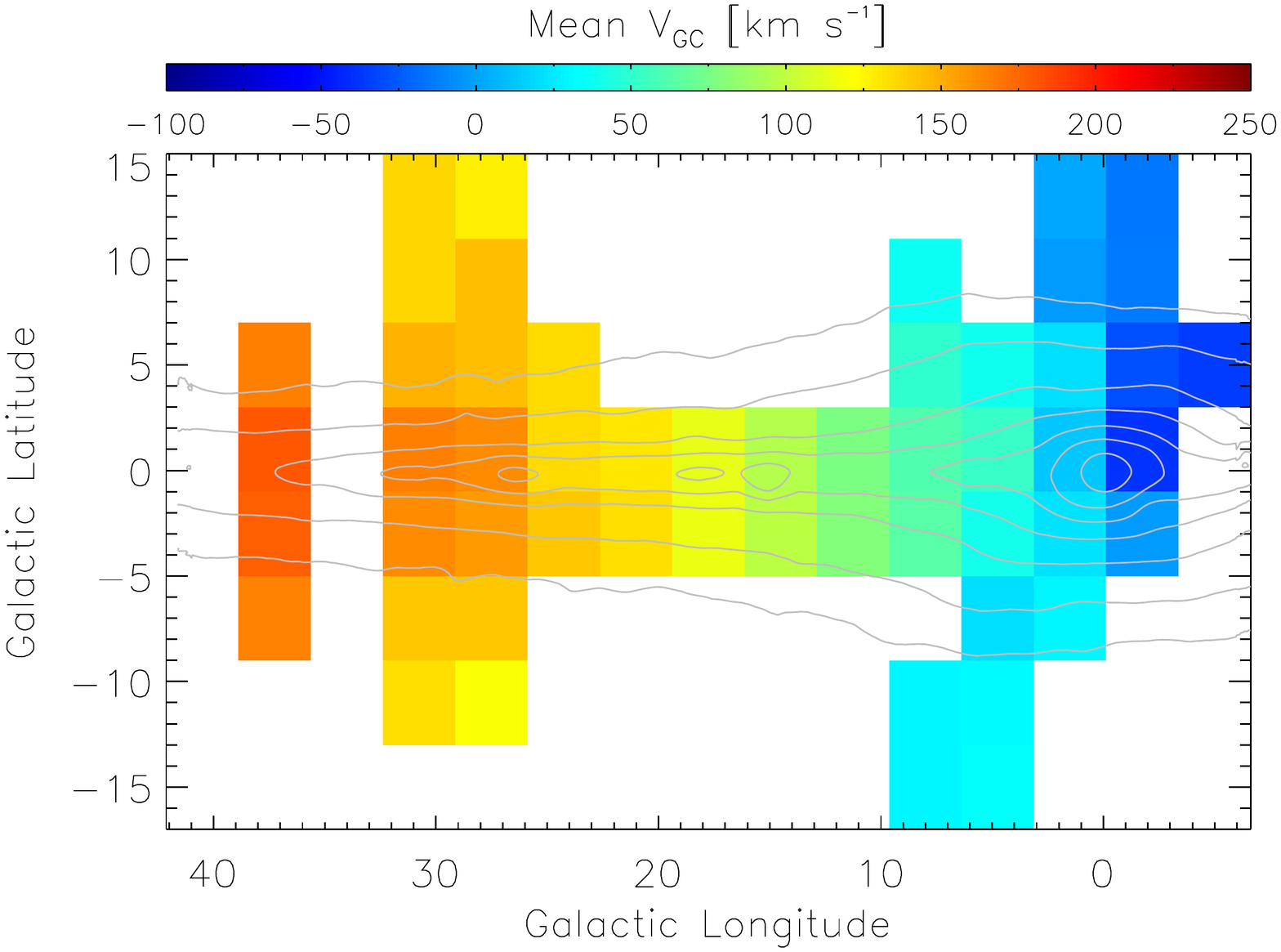} 
  \includegraphics[trim=1in 0.9in 0.5in 0.7in, clip, angle=180, width=0.45\textwidth]{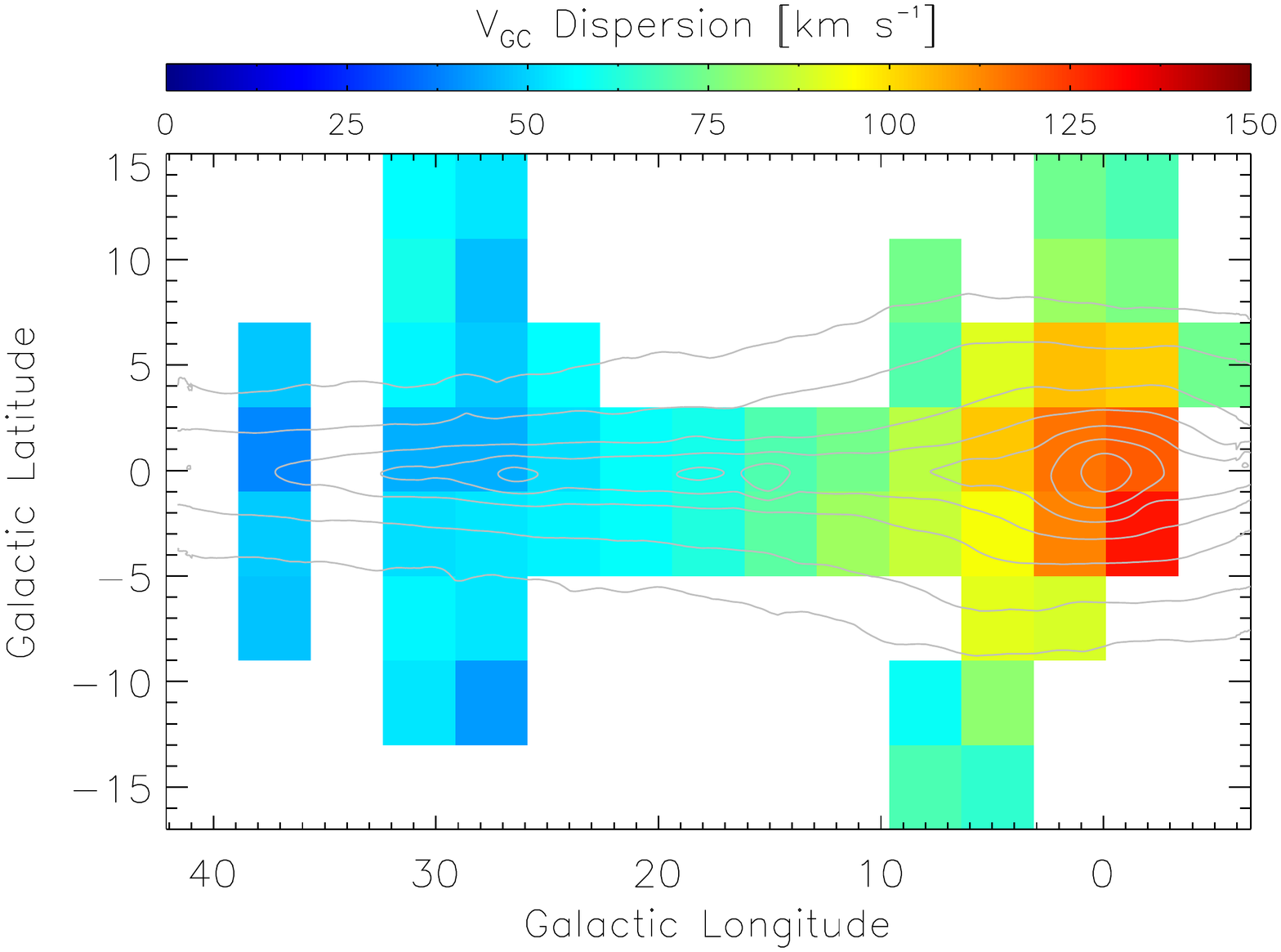} 
\end{center}
\caption{
Maps of velocity mean $<$$V_{\rm GC}$$>$$(l,b)$ and dispersion $\sigma_V(l,b)$ for all stars in our sample.
The gray contours are identical to those in Figure~\ref{fig:WISE_datamap}. 
The stars show cylindrical rotation in the inner Galaxy (see also Section~\ref{sec:cylindrical_rotation}), with a dispersion that peaks at $(l,b) = (0^\circ,0^\circ)$.
}
\label{fig:map_allstars_mean_sig_VGC}
\end{figure*}

\begin{figure*}[]
\begin{center}
  \includegraphics[trim=1in 0.9in 0.5in 0.7in, clip, angle=180, width=0.45\textwidth]{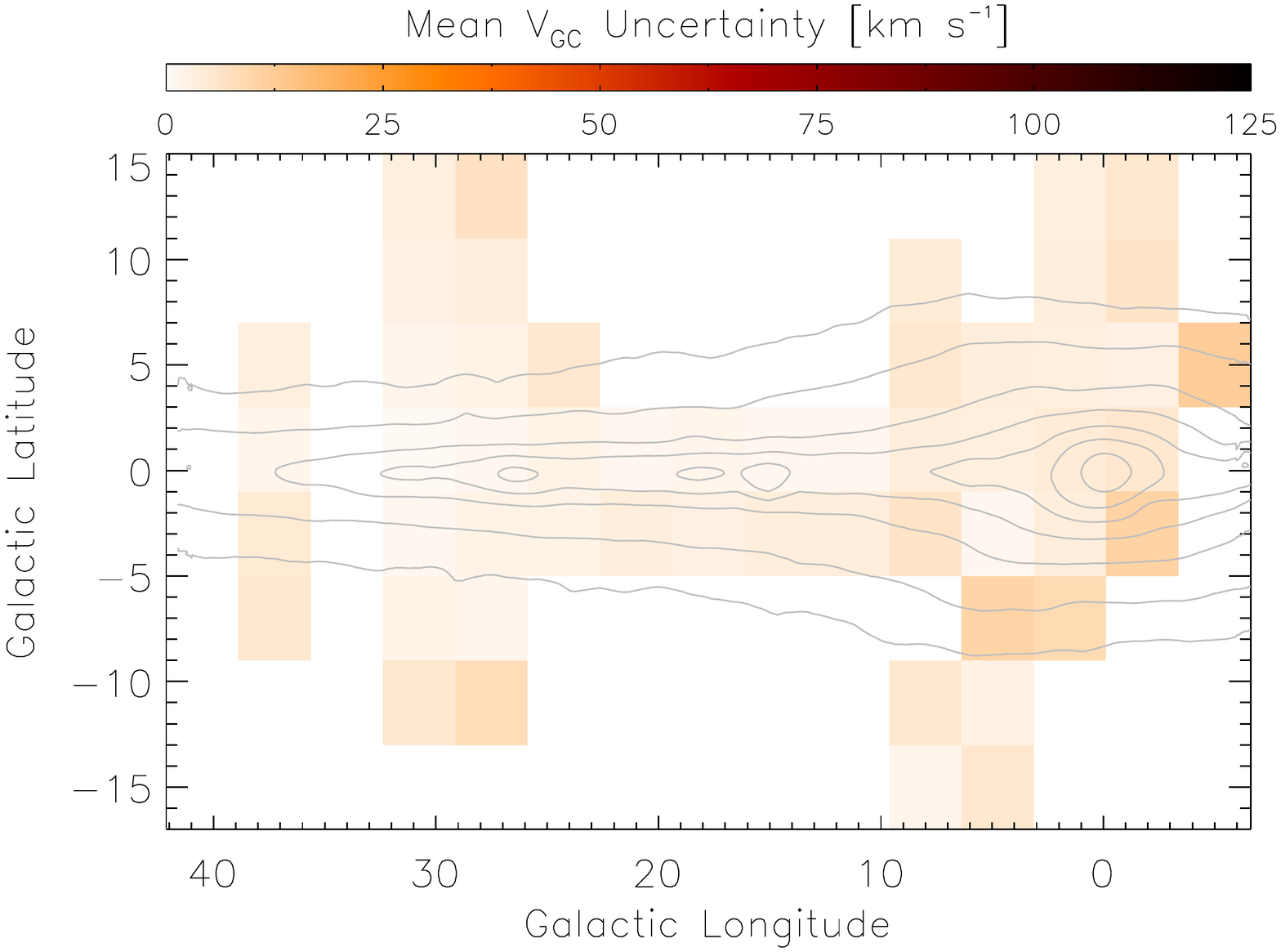} 
  \includegraphics[trim=1in 0.9in 0.5in 0.7in, clip, angle=180, width=0.45\textwidth]{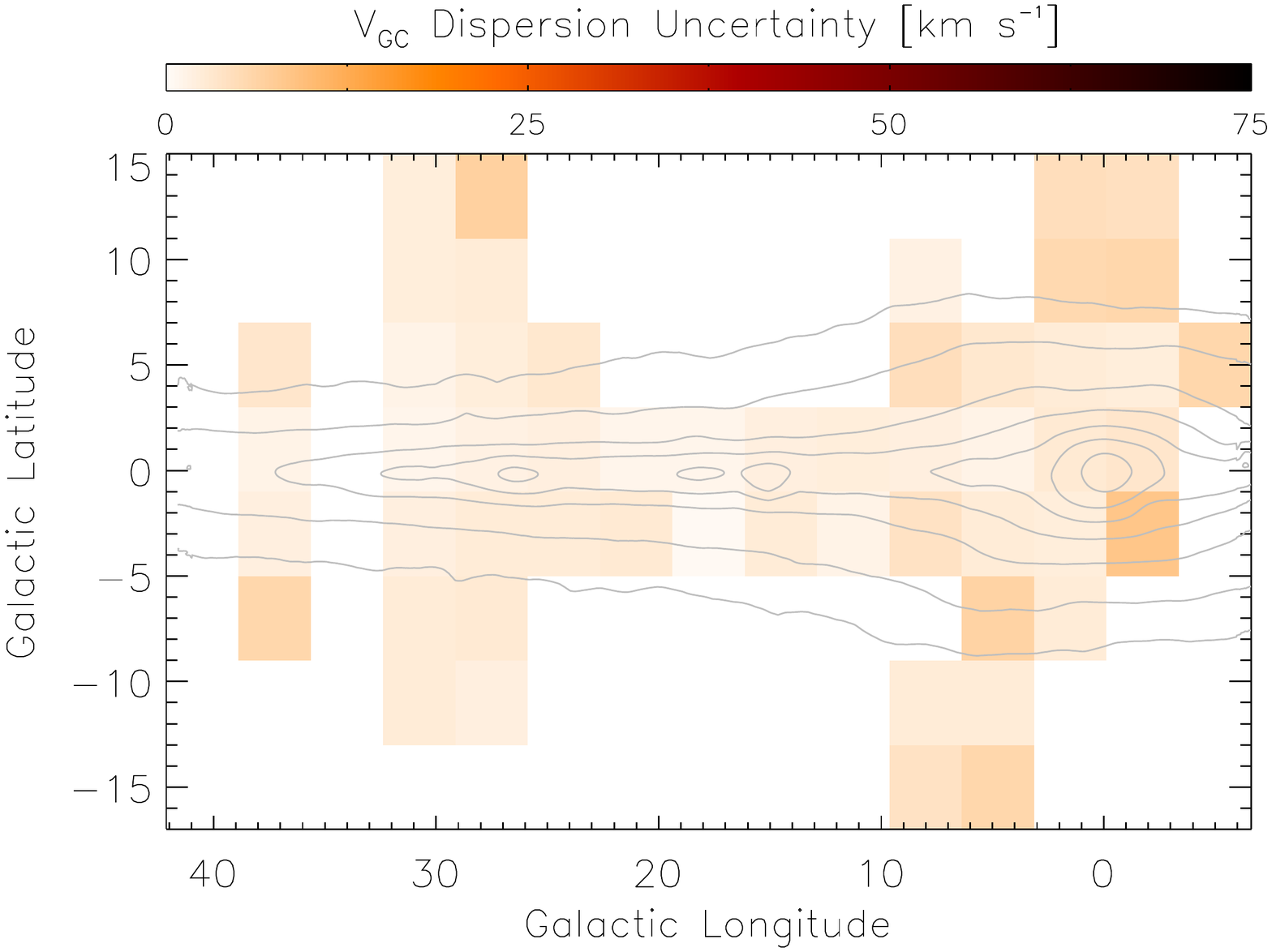} 
\end{center}
\caption{
Bootstrapped uncertainties of the $V_{\rm GC}$ mean and dispersion maps in Figure~\ref{fig:map_allstars_mean_sig_VGC}.
The gray contours are identical to those in Figure~\ref{fig:WISE_datamap}.
Typical mean uncertainties are $\sim$4~km~s$^{-1}$, and typical dispersion uncertainties are $\sim$3~km~s$^{-1}$.
}
\label{fig:map_allstars_mean_sig_VGC_err}
\end{figure*}

In Paper I, we presented maps of the mean rotation velocity, $<$$V_{\rm GC}(l,b)$$>$, and the velocity dispersion, $\sigma(V_{\rm GC})(l,b)$,
for the APOGEE sample in total and for the sample divided into four broad metallicity bins.  
This was the first time these global kinematical patterns had been mapped 
homogeneously and continuously across such a large fraction of the inner Galaxy ($\sim$550~deg$^2$).
Reproductions of these patterns for the total sample
can be seen in Figure~\ref{fig:map_allstars_mean_sig_VGC} (and their uncertainties in Figure~\ref{fig:map_allstars_mean_sig_VGC_err});
these include a smooth increase in the mean velocity with longitude, little dependence
of the mean velocity on latitude, and a velocity dispersion that peaks towards the Galactic Center 
and rapidly decreases with both longitude and latitude.
The rotation and dispersion patterns in the (previously unexplored) midplane and inner bulge are consistent with those observed
at higher latitudes.
We note that the offset of the peak of the dispersion from $(l,b)=(0^\circ,0^\circ)$ 
is simply a result of the angular bin sizes in Figure~\ref{fig:map_allstars_mean_sig_VGC}, 
chosen to yield enough stars for a significant measurement of the higher order moments (Section~\ref{sec:high_moments}),
and the spacing of the stars within those bins.

\begin{figure}[]
\begin{center}
  \includegraphics[trim=1in 0.9in 0.5in 0.7in, clip, angle=180, width=0.5\textwidth]{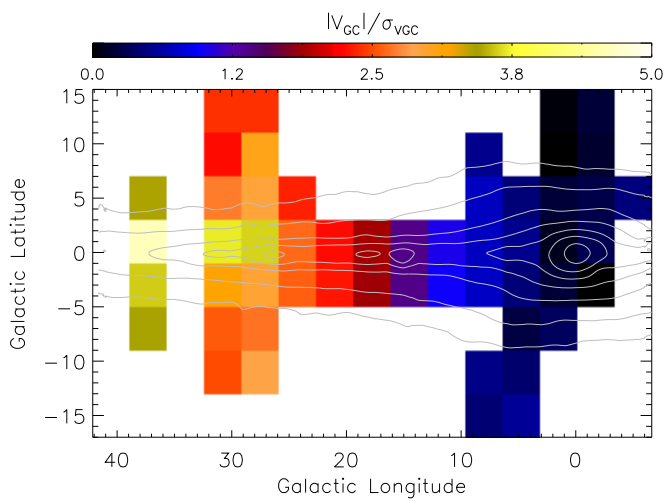} 
\end{center}
\caption{
$|V_{\rm GC}|/\sigma_V$ for all stars.
The gray contours are identical to those in Figure~\ref{fig:WISE_datamap}.
}
\label{fig:map_allstars_mean_over_sig_VGC}
\end{figure}

In Figure~\ref{fig:map_allstars_mean_over_sig_VGC}, we show a new map of $|$$<$$V$$>$$|/\sigma_V$ for these stars.  
We find $|$$<$$V$$>$$|/\sigma_V<1$ in the inner bulge ($l \lesssim 10^\circ$) at all latitudes, 
increasing to $|$$<$$V$$>$$|/\sigma_V \sim 4$ in the rotationally-supported inner disk.

We note how this type of dataset is (and is not) approaching an ``integrated light view'' of the MW, 
such as we have in external edge-on galaxies.
On one hand, the increased stellar sample size and coverage (especially with IR surveys like APOGEE) 
afford the opportunity to make smooth, nearly continuous measurements of the MW's
mean kinematical properties, instead of relying on fields (often centered on low extinction windows)
that are small relative to the angular distances being interpolated between them.
This produces maps that are, at least visually, comparable to those extracted 
from long-slit or IFU measurements of external systems.

On the other hand, our location within the MW produces some perspective effects not encountered in sightlines through distant galaxies,
such as the ``cone'' effect, in which angular bins defined in $l$ and $b$ span a greater range of volume at larger distances.
Furthermore, stellar datasets in the MW are generally magnitude limited (affected by both extinction and distance), 
whereas the unresolved stellar contributions to extragalactic light are flux limited (susceptible to extinction but not individual distance effects).
Kinematical maps of the inner MW are most often number count weighted, and do not include contributions from stars
towards the anti-center, which would appear in inner MW sightlines seen from an external vantage point.
These anti-center stars are few in number, compared to the inner Galaxy, but may have systematic influences
in galaxies seen edge-on.
The velocity field measured in a true integrated light spectrum results from {\it all} stars along the line of sight and 
is affected by complex weightings that are dependent
on, e.g., the stellar luminosity, stellar metallicity, and wavelength of the observation.  
These weightings can be calculated for MW datasets, but a full treatment requires a number of other assumptions
and is reserved for future work.

\subsection{Velocity Skewness and Kurtosis} \label{sec:high_moments}

\begin{figure*}[]
\begin{center}
  \includegraphics[trim=1in 0.9in 0.5in 0.7in, clip, angle=180, width=0.45\textwidth]{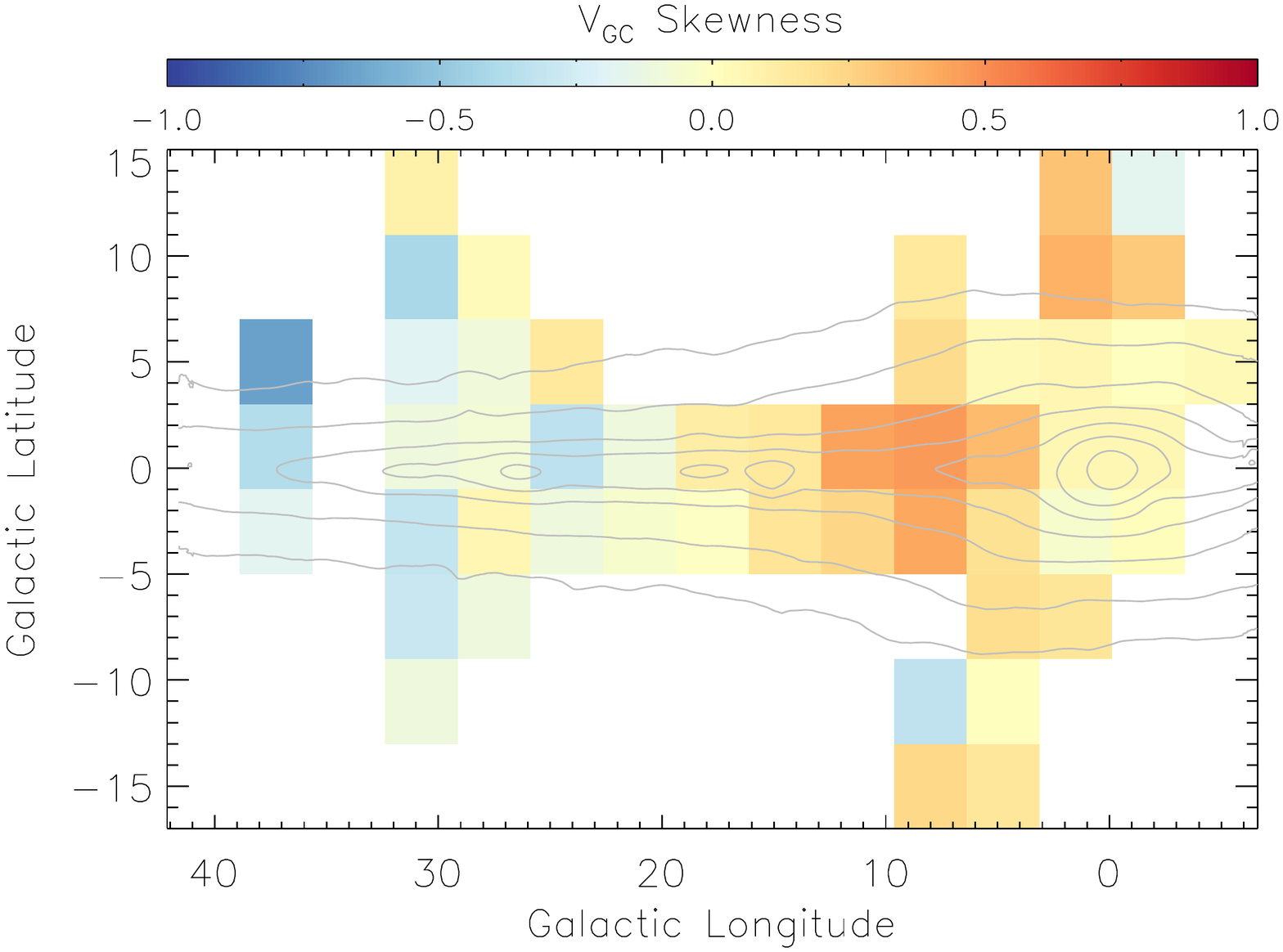} 
  \includegraphics[trim=1in 0.9in 0.5in 0.7in, clip, angle=180, width=0.45\textwidth]{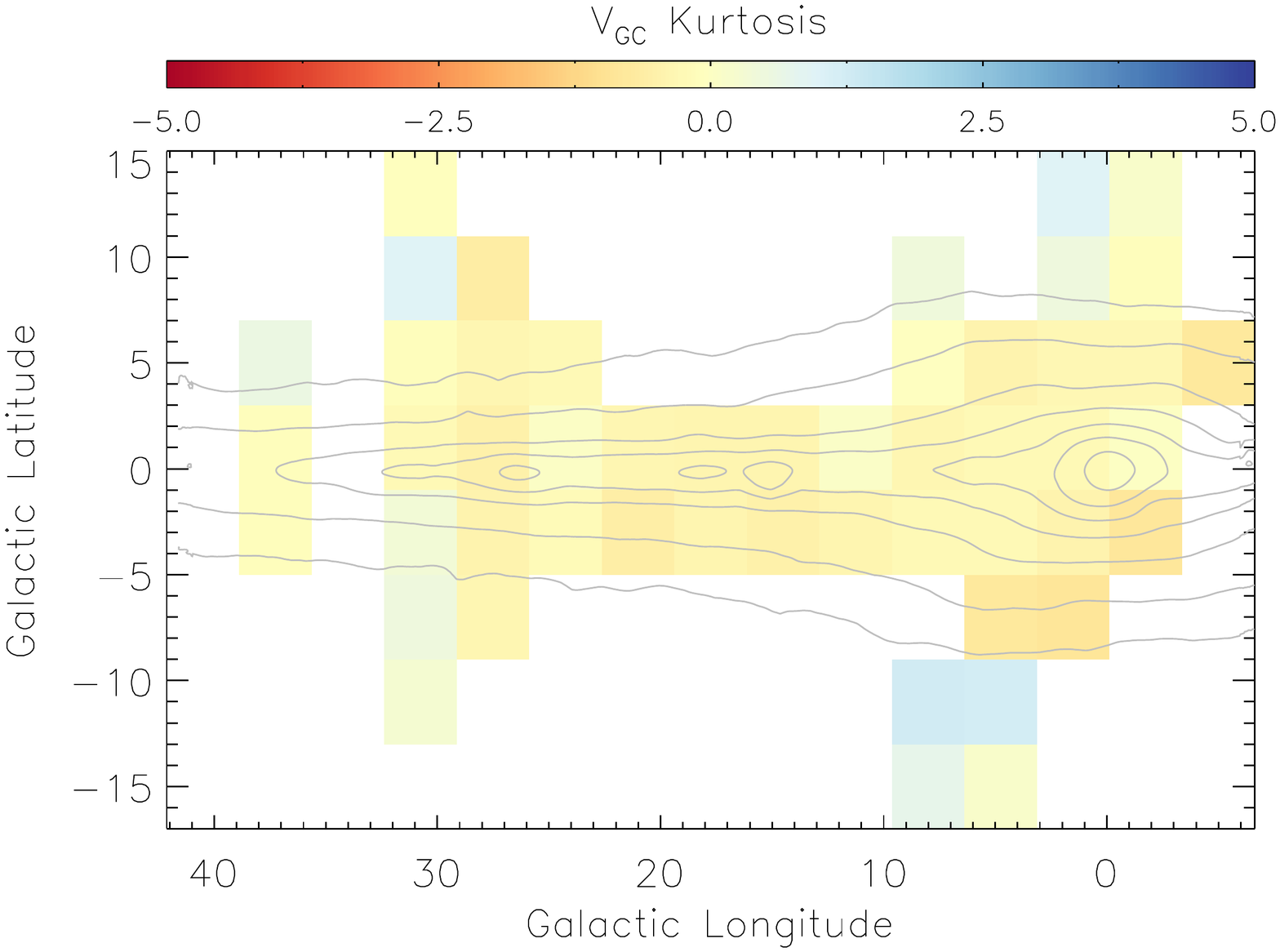} 
\end{center}
\caption{
Maps of $V_{\rm GC}$ skewness and kurtosis for all stars in our sample.
The gray contours are identical to those in Figure~\ref{fig:WISE_datamap}.
The LOSVDs are positively skewed in the midplane around $l \sim 10^\circ$, 
and the kurtosis is roughly zero in nearly all bins within the uncertainties (Figure~\ref{fig:map_allstars_h3_h4_VGC_err}).
}
\label{fig:map_allstars_h3_h4_VGC}
\end{figure*}

\begin{figure*}[]
\begin{center}
  \includegraphics[trim=1in 0.9in 0.5in 0.7in, clip, angle=180, width=0.45\textwidth]{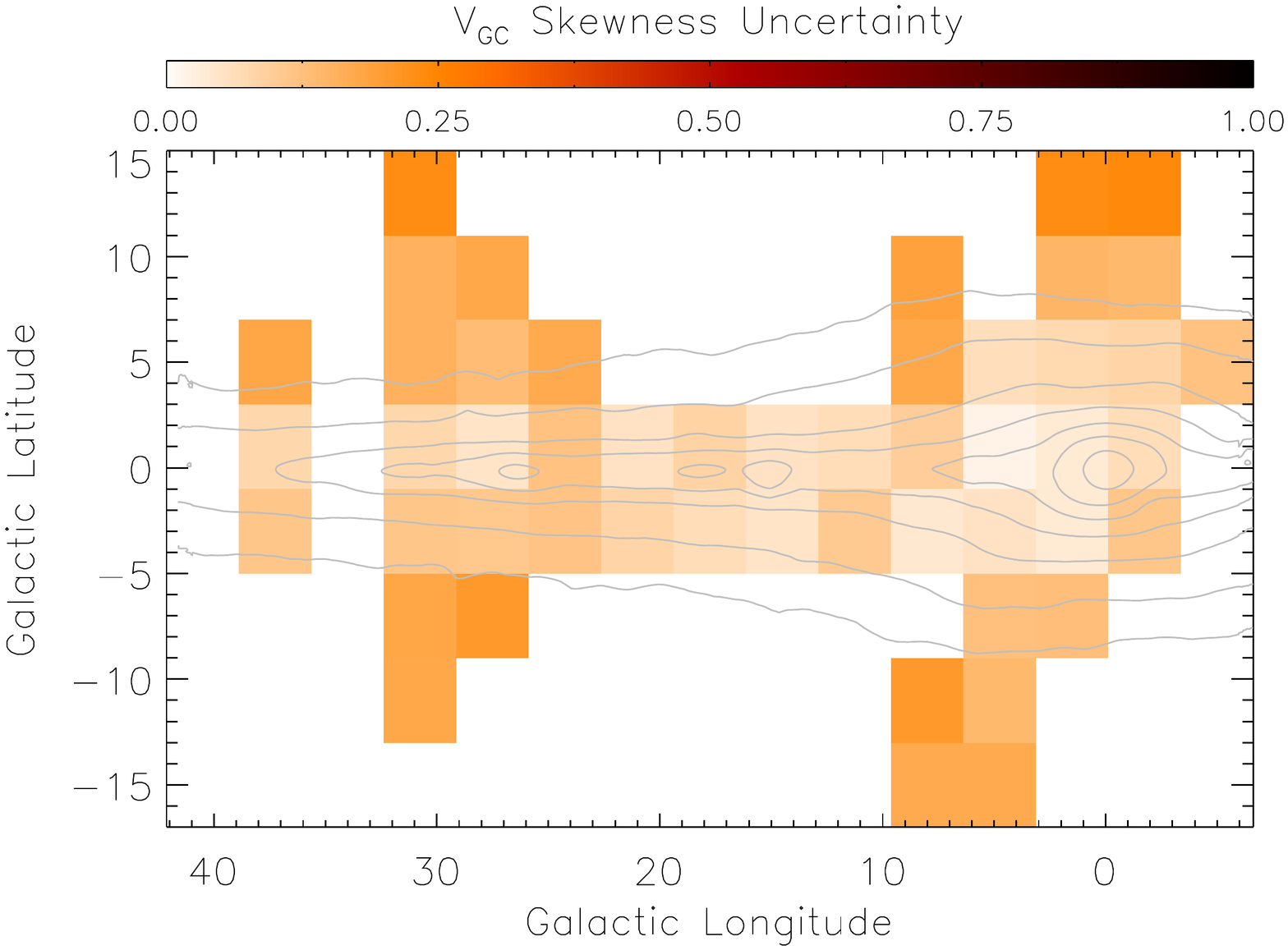} 
  \includegraphics[trim=1in 0.9in 0.5in 0.7in, clip, angle=180, width=0.45\textwidth]{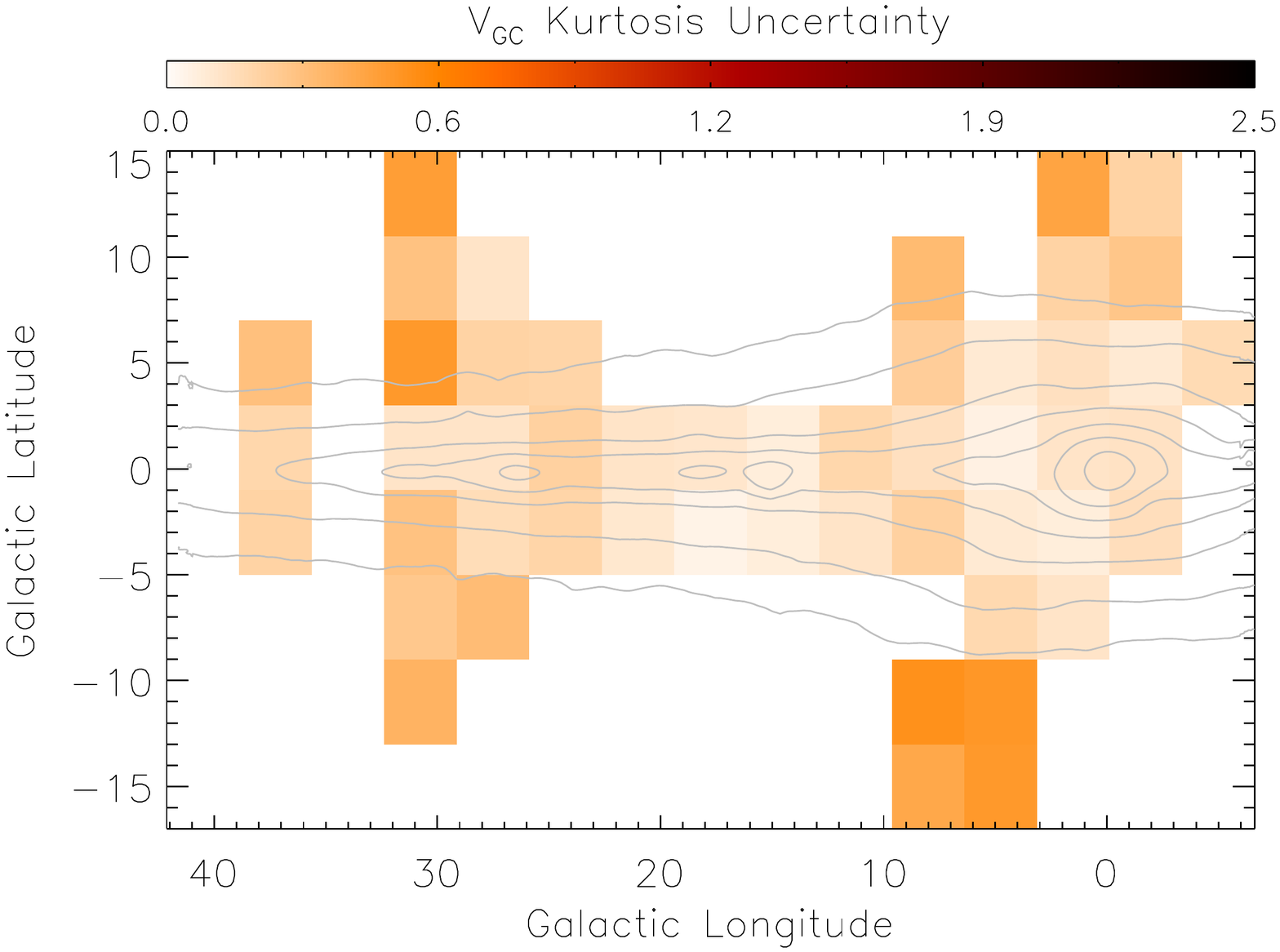} 
\end{center}
\caption{
Bootstrapped uncertainties of the $V_{\rm GC}$ skewness and kurtosis maps in Figure~\ref{fig:map_allstars_h3_h4_VGC}. 
The gray contours are identical to those in Figure~\ref{fig:WISE_datamap}.
Typical skewness uncertainties are $\sim$0.1, and typical kurtosis uncertainties are $\sim$0.2.
}
\label{fig:map_allstars_h3_h4_VGC_err}
\end{figure*}

The $<$$V$$>$ and $\sigma(V)$ maps show clear patterns that are generic 
features of many barred galaxy models (Paper I and Section~\ref{sec:model_data}),
including both cosmologically motivated simulations and $N$-body models evolved from a disk of particles.
These simulations display similar global rotation and dispersion trends, 
despite different initial conditions, star formation histories, and final bar aspect ratios and masses.
Because the models are scaled to certain properties of the MW, 
and sometimes tuned to match the observational $<$$V$$>$ and $\sigma(V)$ patterns from bulge surveys,
the higher order moments of the LOSVDs offer stronger observational constraints to examine the consequences
of the models' unique initial conditions and detailed evolutionary histories.

Determining these moments observationally, to use as constraints on models, requires a large, relatively unbiased sample of stars.
In Figure~\ref{fig:map_allstars_h3_h4_VGC}, we present the first
maps of the third and fourth moments (skewness and kurtosis) of stars in the MW bulge.
The uncertainties are shown in Figure~\ref{fig:map_allstars_h3_h4_VGC_err}.

\begin{figure}[!hptb]
\begin{center}
  \includegraphics[trim=1.0in 2.0in 4.2in 1.1in, clip, width=0.4\textwidth]{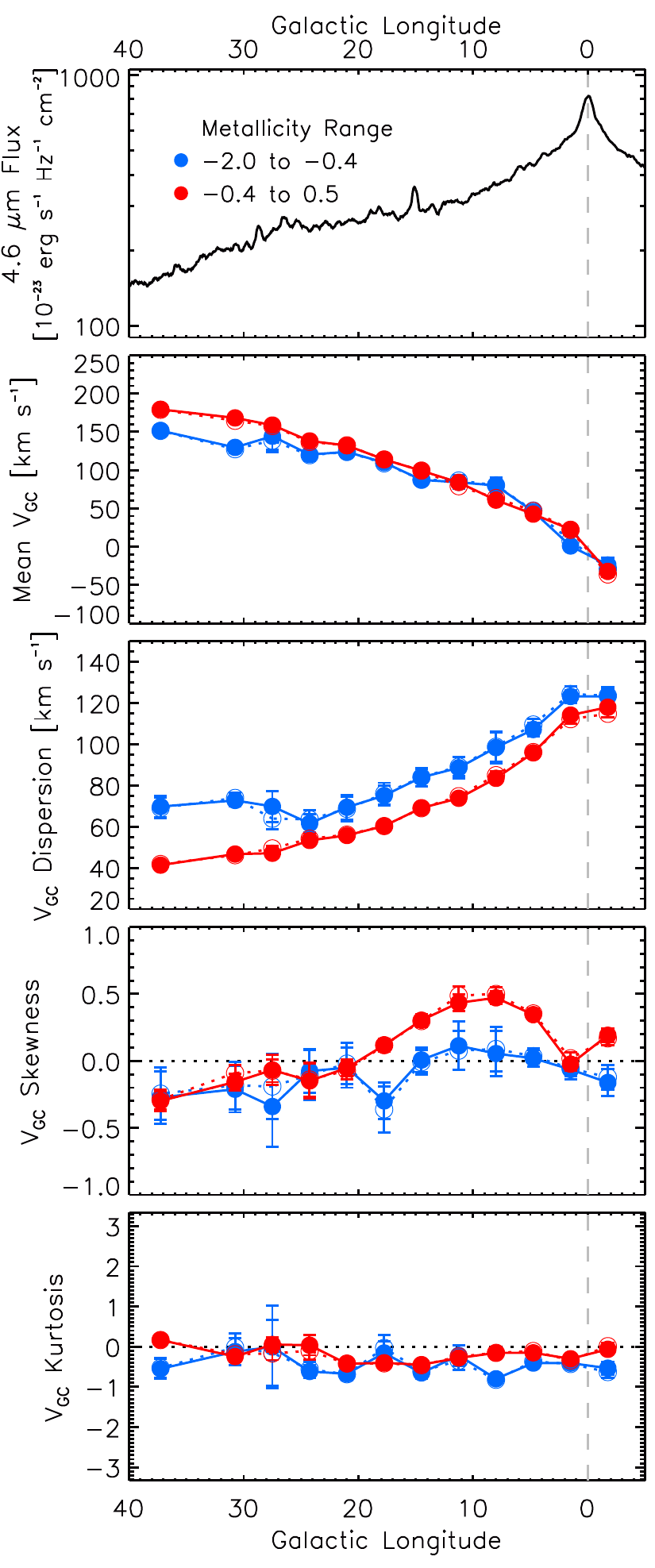} 
\end{center}
\caption{
Midplane ($|b|<5^\circ$) velocity moments for two metallicity bins: ``metal-poor'' (blue, ${\rm [Fe/H]} < -0.4$) and ``metal-rich'' (red, ${\rm [Fe/H]} \ge -0.4$).  
Uncertainties are derived from a bootstrap resampling of the data.  
Dotted lines indicate values derived from the raw counts, and the solid lines have been corrected for the APOGEE selection function.  
The dashed gray line indicates $l=0^\circ$.
Note the similar mean rotation and kurtosis behavior, but dissimilar dispersion and skewness patterns, between the metal poor and metal rich stars.
}
\label{fig:moments_plots}
\end{figure}

These maps, which contain all of the stars in our sample, begin to approach an ``integrated light'' view of the inner Galaxy, 
modulo the issues discussed in Section~\ref{sec:low_moments}.
In Figure~\ref{fig:moments_plots}, we show the 1-D kinematical
patterns against Galactic longitude, for all stars with $|b| \le 5^\circ$.  This is conceptually similar to placing a spectroscopic slit
along the major axis of an edge-on bulge/bar and measuring the velocity parameters at multiple positions along the slit.
However, one critical difference --- and the reason why the MW is such a powerful tool for this kind of work --- is that we can
perform these analyses independently for stars of different chemistry.  

Thus, in Figure~\ref{fig:moments_plots}, the red points indicate the 1-D kinematical behavior for more metal rich stars
(${\rm [Fe/H]} > -0.4$), and the blue points indicate the same for more metal poor stars ($-2 \le {\rm [Fe/H]} \le -0.4$).\footnote{The ${\rm [Fe/H]} \ge -2$ 
requirement removes a tiny fraction ($<$0.1\%) of stars for which the {\it Cannon} assigns 
values inconsistent with the stars' visually inspected spectra, in part because of inadequate labelspace coverage.}
The dotted lines and open circles indicate the mean velocity, dispersion, skewness, and kurtosis derived
from the raw, count-weighted velocity KDEs, and the solid lines and circles show these values derived from
velocity KDEs corrected for the APOGEE selection function.\footnote{The selection function corrections are derived from the 
comparison of candidate target stars to the observed stars, in each field-color-magnitude bin.  Details of the color and magnitude bins
used for defining the APOGEE sample can be found in \citet{Zasowski_2013_apogeetargeting}.}

In Paper I, we adopted four metallicity bins to highlight the gradual changes in $<$$V$$>$$(l,b)$ and $\sigma(V)$$(l,b)$
with chemistry; here, we use only two bins in order to make statistically meaningful measurements of ${\rm Skew}(V)$$(l)$ and ${\rm Kurt}(V)$$(l)$,
which require more stars than measuring the mean and dispersion.
The dividing value of ${\rm [Fe/H]} = -0.4$ was chosen as being approximately where the kinematical behavior began to change amongst
the four bins in Paper I --- between bins B and C in that paper, at ${\rm [Fe/H]} = -0.5$.
Other values within the range $-0.8 < {\rm [Fe/H]} < -0.3$ were tested, 
and splitting the sample at ${\rm [Fe/H]} = -0.4$ was found to produce the same qualitative results
as lower metallicities but with better statistics and correspondingly smaller uncertainties.

The key points of Figure~\ref{fig:moments_plots} can be summarized as follows:
\vspace{-8pt}
\begin{itemize} \itemsep -2pt
\item The mean velocities are very consistent between the two groups at all longitudes.  
(This consistency is in agreement with Paper I; recall that Paper I's finding of slower rotation among metal-poor stars was most significant 
in the most metal poor bin there, with ${\rm [Fe/H]} \le -1$.)
The velocity dispersions have similar qualitative patterns (peak towards the Galactic Center, decline into the disk), 
but the lower-[Fe/H] sample has a higher dispersion overall, and this difference increases at higher longitudes.  
This may indicate an increasing proportion of thick disk stars in this group.
\item We see an indication of negative LOSVD skewness (a low velocity tail) towards the end 
of the long bar ($l \sim 30-35^\circ$), in both metallicity bins within the uncertainties.  
However, we note that this behavior may be ascribed to the presence of a few very low-velocity outliers 
(slightly visible in the histograms in Figure~\ref{fig:RV_histograms}, e.g., at $[l,b] = [30^\circ, -2^\circ]$).  
Without these outliers, the LOSVDs at these longitudes are extremely symmetric.
\item By far the most dramatic feature seen in the skewness is the high ${\rm Skew}(V)$ peak centered at $l \sim 10^\circ$, 
decreasing smoothly over several degrees to either side.  Remarkably, this pattern appears only in the high-[Fe/H] sample, 
to a very high level of confidence.  This signal is a more robust way of quantifying the high velocity ``peaks'' first
published in \citet{Nidever_2012_apogeebar} and further analyzed in, 
e.g., \citet{Aumer_2015_highRVbarstars} and \citet{Debattista_2015_kpcnucleardisk}.
We discuss this result in light of the high-RV ``peaks'' in Section~\ref{sec:high_velocity} and in the context of
extragalactic bar diagnostics in Section~\ref{sec:extragal}.
\item We do not find significant kurtosis in any of the longitude bins.
\end{itemize}

\begin{figure}[!hptb]
\begin{center}
  \includegraphics[trim=2.5in 1.1in 1.2in 5.2in, clip, angle=180, width=\textwidth]{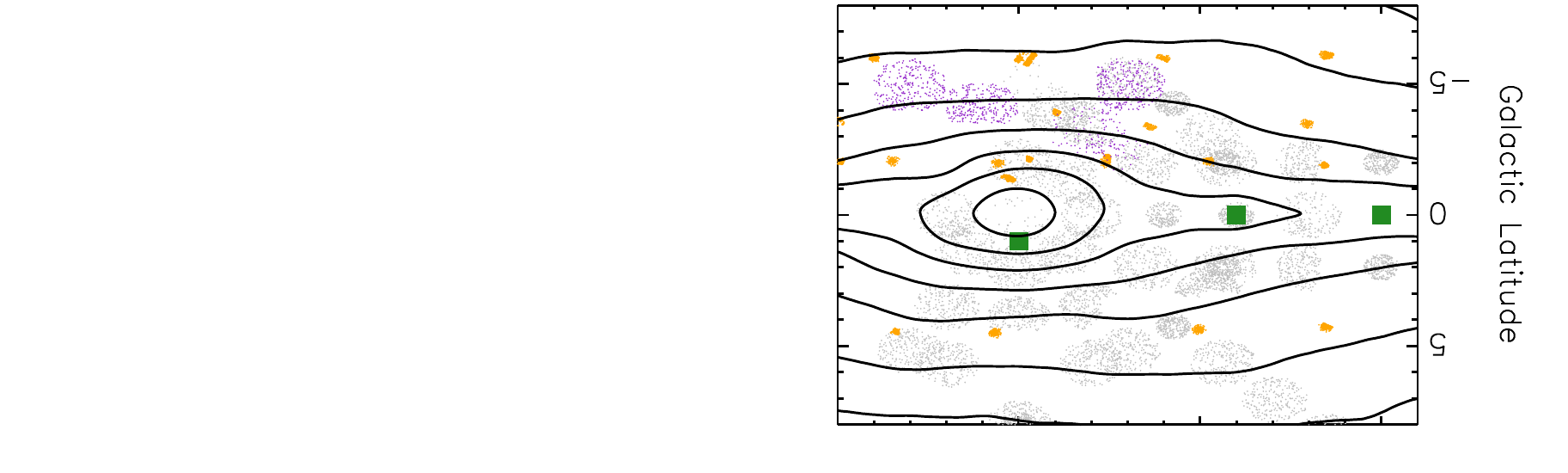} 
  \includegraphics[trim=2.5in 1.3in 1.2in 1.3in, clip, angle=180, width=\textwidth]{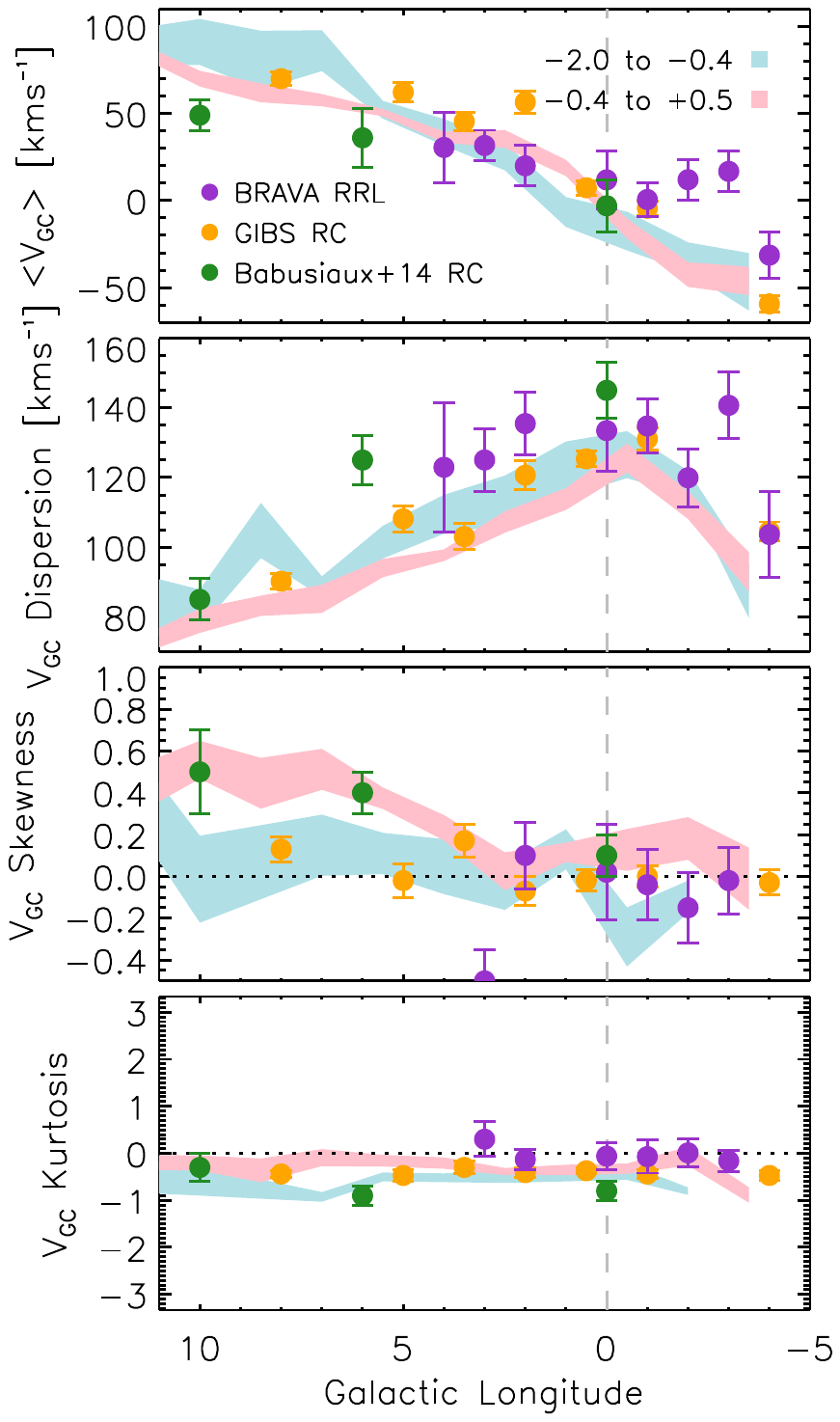} 
\end{center}
\caption{
Midplane velocity moments for the APOGEE samples as in Figure~\ref{fig:moments_plots} ($\pm$1$\sigma$ trends in pink and light blue), 
compared to those of the RRL sample from BRAVA \citep[][purple]{Kunder_2016_bulgeRRLbrava}, and
of the RC samples from \citet[][green]{Babusiaux_2014_barmetallicitykinematics} 
and the GIBS survey \citep[][orange]{Zoccali_2014_GIBS1}.  
The top panel shows the distribution of the APOGEE points (gray) and these comparison samples (colored), 
along with the WISE flux contours \citep{Lang_2014_unWISEcoadds}.
The mean rotation, dispersion, and kurtosis are consistent among the non-RRL samples, 
but the high skewness pattern is seen only in APOGEE and the Babusiaux RC stars, which probe closest to the midplane.
}
\label{fig:compare_moments_plots}
\end{figure}

In Figure~\ref{fig:compare_moments_plots}, we compare the behavior of these moments in the innermost degrees with those
of three other observed datasets probing close to the Galactic midplane (restricted to $|b| \le 5^\circ$): two samples of red clump (RC) stars from 
\citet{Babusiaux_2014_barmetallicitykinematics} and the GIBS survey \citep{Zoccali_2014_GIBS1},
and the set of RR Lyrae (RRL) stars observed by BRAVA \citep{Kunder_2016_bulgeRRLbrava}.  
The latitude distributions of these samples are not perfectly matched to the APOGEE coverage, 
which may explain some of the discrepancies, but overall we find good agreement, or differences where expected.
The mean velocities are roughly in agreement with the exception of the RRL sample (particularly noticeable at $l < 0^\circ$),
whose longitude-independent velocities suggest a nearly non-rotating population \citep{Kunder_2016_bulgeRRLbrava}.
Both the RRL and the \citet{Babusiaux_2014_barmetallicitykinematics} RC stars have a higher dispersion than the APOGEE sample,
and this high dispersion was noted in comparison to other studies in those papers as well.

Only the RC samples extend beyond $l \sim 5^\circ$, where the LOSVDs of the metal-rich APOGEE stars show the large positive skewness.
This positive skewness is seen very clearly in the \citet{Babusiaux_2014_barmetallicitykinematics} RC stars as well
(and the presence of high velocity tails was noted in that work), but it is not observed in the GIBS sample.
This highlights the concentration of the high velocity stars towards the lowest latitudes, 
which are represented in the APOGEE and Babusiaux RC stars but not in the GIBS RC stars 
(where $|b| \gtrsim 2^\circ$; see also Section~\ref{sec:high_velocity}).
None of the comparison samples show significant kurtosis, in agreement with the APOGEE data.

\begin{figure}[!hptb]
\begin{center}
  \includegraphics[trim=4.5in 6.3in 0in 1.8in, clip]{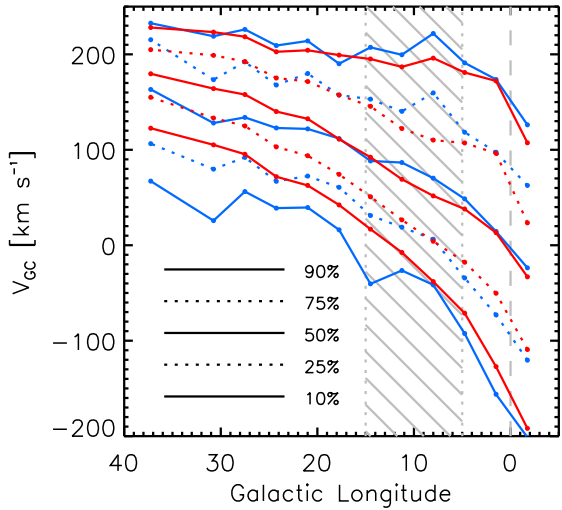} 
\end{center}
\caption{
Quantiles (10\%, 25\%, 50\%, 75\%, 90\%) of the midplane velocity distributions, using the same longitude binning, latitude limit, and color scheme as Figure~\ref{fig:moments_plots}.  The dashed gray line indicates $l=0^\circ$, and the gray cross-hatched region indicates $l=5-15^\circ$, 
where the difference in LOSVD skewness between the metal-rich and metal-poor samples is largest.
}
\label{fig:quantiles}
\end{figure}

In Figure~\ref{fig:quantiles}, we show an alternate representation of the LOSVD shapes, 
in the form of curves of the 10\%, 25\%, 50\%, 75\%, and 90\% quantiles of the distributions.
As in Figure~\ref{fig:moments_plots}, blue lines indicate properties of the metal-poor group, 
and red lines indicate the metal-rich one.
The impact of the metal-poor, low velocity outliers described above can be seen in the blue 10\% curve,
and the skewness differences between the populations is visible in the differences in the 75\% and 90\% curves around $l \sim 10^\circ$.
For the remainder of this paper, however, we will use the LOSVD moments to describe the kinematical characteristics.

\subsection{Comparison to N-Body Models} \label{sec:model_data}

\begin{figure*}[!hptb]
\begin{center}
  \includegraphics[trim=1in 1in 0.5in 2.1in, clip, angle=180, width=0.49\textwidth]{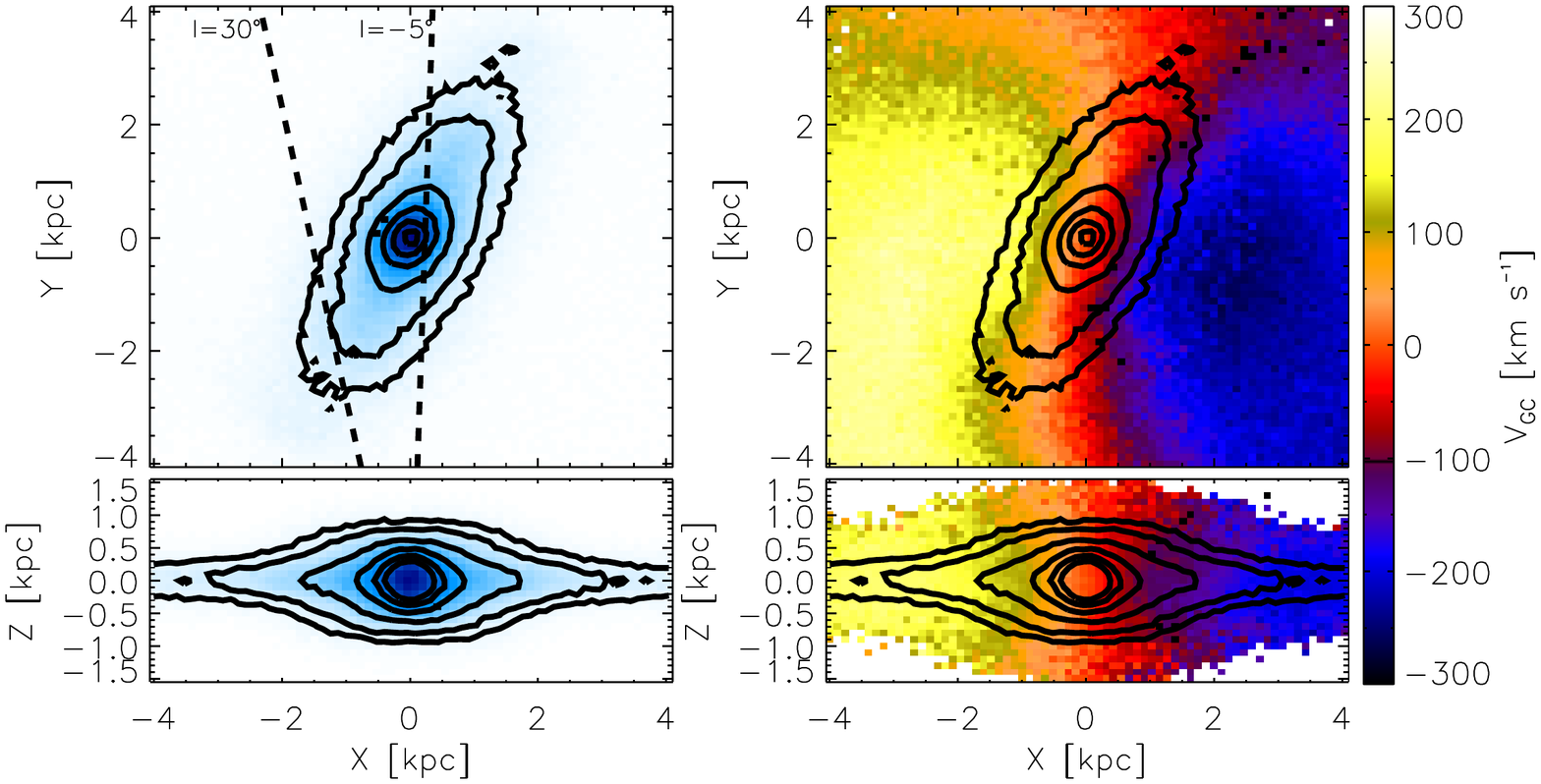} 
  \includegraphics[trim=1in 1in 0.5in 2.1in, clip, angle=180, width=0.49\textwidth]{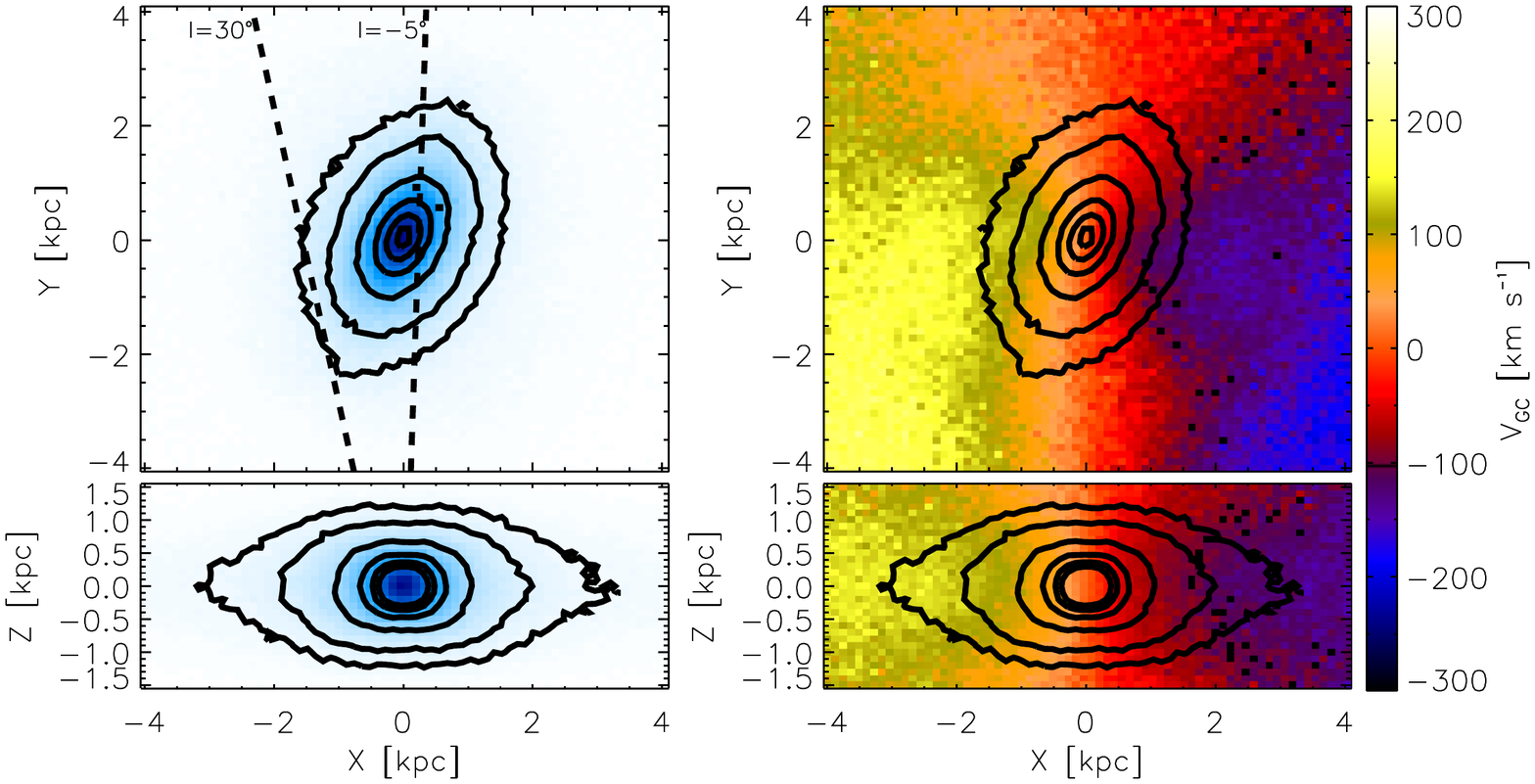} 
\end{center}
\caption{
Particle density (bluescale) and radial velocity (colored) distributions for the MW-scaled $N$-body 
models of \citet[][left]{MartinezValpuesta_11_barmodel}
and \citet[][right]{Shen_10_purediskbulge}, shown in the {\it XY} and {\it XZ} planes.
The contours indicate the particle density (5\%, 10\%, 25\%, 50\%, 75\%, and 95\% of maximum) 
and are repeated in the corresponding velocity maps.
Velocities are $V_{\rm GC}$ from the position of the Sun ($X=0$~kpc, $Y=-8$~kpc).
The dashed lines indicate the positions of the $l=30^\circ$ and $l=-5^\circ$ sightlines.
}
\label{fig:models}
\end{figure*}

We compare our data to two {\it N}-body barred galaxy models (particles only, without gas or star formation) 
that are evolved from an initially thin Galactic disk. 
This disk grows a bar, which experiences vertical buckling instabilities and forms a boxy/peanut bulge at early times
\citep[e.g.,][]{Combes_1981_barformation,Raha_1991_bars}. 
These particular models are discussed in more detail in \citet[][hereafter MVG]{MartinezValpuesta_11_barmodel} and in 
\citet[][hereafter S10]{Shen_10_purediskbulge}.
The modeled bars are qualitatively similar in morphology to that observed in the MW,
and are scaled to approximate the bar's half-length and the circular velocity at the position of the Sun.
Both have been observed to approximately reproduce the MW's behavior in their velocity means and dispersions, 
so it is interesting to see here which higher order velocity properties are generic features of bars 
and which properties have a strong dependence on the choice of model.

The spatial densities and velocity fields for the MVG model are shown in Figure~\ref{fig:models}a, 
and for the S10 model in Figure~\ref{fig:models}b.
For both, the blue shading and black contours in the left panels indicate the spatial density of particles in the {\it XY} (top) and {\it XZ} (bottom) planes, 
with the central bar's major axis angled 27$^\circ$ from the Sun-Galactic Center line.  
The dashed lines in the {\it XY} planes mark the sightlines at $l=-5^\circ$ and $l=30^\circ$, 
the approximate edges of the APOGEE sample used here.
The rainbow panels have identical particle density contours and viewing perspectives; the pixel colors
indicate the mean Galactocentric velocity (relative to the Sun's position at $X=0, Y=-8$) of particles in each pixel.

\begin{figure}[!hptb]
\begin{center}
  \includegraphics[trim=1.0in 2.0in 4.2in 2.85in, clip, width=0.4\textwidth]{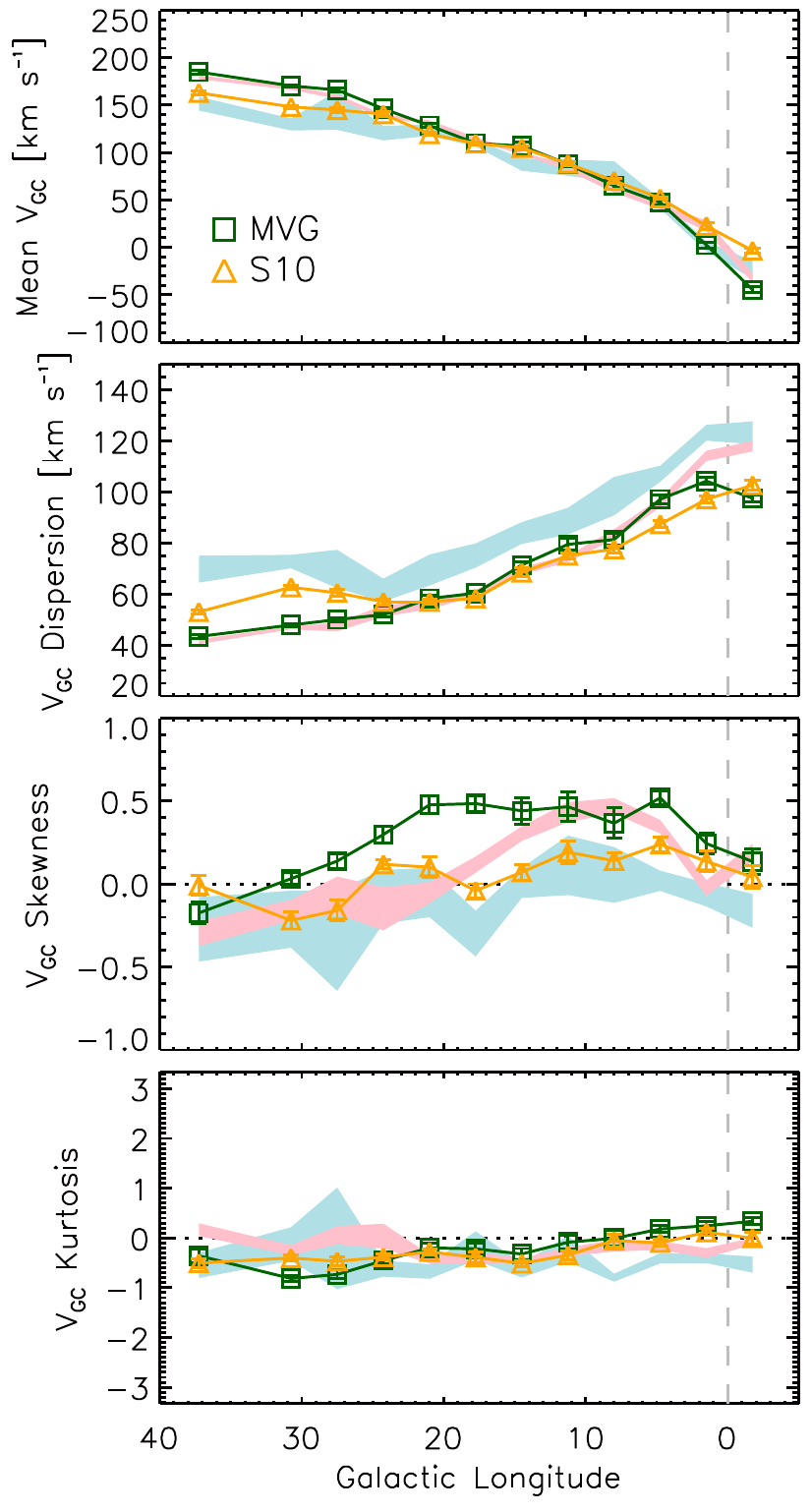} 
\end{center}
\caption{
Similar to Figure~\ref{fig:moments_plots}, where the pale blue and red shadings here are the $\pm$1$\sigma$ blue and red trends from that figure.
Overplotted are the comparable kinematical properties from the \citet[][MVG]{MartinezValpuesta_11_barmodel} and
\citet[][S10]{Shen_10_purediskbulge} models, as described in the text.
Good agreement is seen in the mean rotation and kurtosis, but both models underpredict the observed central dispersion,
and the skewness patterns are qualitatively different from that of the dominant metal rich population.
}
\label{fig:moments_plots_models}
\end{figure}

Figure~\ref{fig:moments_plots_models} shows the 1-D kinematical patterns for these models, 
using green squares for MVG and orange triangles for S10.  
As in Figure~\ref{fig:compare_moments_plots}, 
the observed APOGEE moments ($\pm 1\sigma$) from Figure~\ref{fig:moments_plots} are included as lighter red and blue shading in the background.
For this comparison, the models have been sampled at the approximate 3D positions of the APOGEE stars, 
to avoid biases due to mismatches between the model density and the APOGEE sampling of the MW.
Because no stellar observables are included in the simulations, we could not apply the actual APOGEE selection function,
and a more complex model sampling that includes potential APOGEE distance systematics will be addressed in future work.
In the current scheme, only 5\% of the APOGEE stars do not have nearby particles in the MVG model, and 1\% in the S10 model;
these are predominantly stars within 2~kpc of the Sun, where the models' densities are low.  Removing these stars
from the observational results described in Sections~\ref{sec:low_moments}--\ref{sec:high_moments} and elsewhere in this
paper does not affect the outcomes or our conclusions.

As expected, we find a very good match between the data and the simulations' mean velocities at most longitudes, especially with the MVG model.
The dispersion in both models is $\sim$20~km~s$^{-1}$ lower than in the data towards the center of the Galaxy, 
which was also noted in Paper I in the comparison with yet another $N$-body simulation \citep{Athanassoula_2007_barmodel}, 
though no sampling to match APOGEE's distance distribution was applied there.
We note that if the MVG and S10 models are {\it not} downsampled to match the APOGEE spatial distribution,
their dispersions at $l=0^\circ$ increase to the level observed in APOGEE, largely due to the higher number of $d > 8$~kpc particles being included.
This could suggest that the APOGEE-{\it Cannon}
distances are systematically too short in this direction, or that the models' $R_{\rm GC}-\sigma_V$ behaviors are offset from the data's,
among other possibilities.
Further into the inner disk, both model dispersions match that of 
the metal-rich stars --- the dominant population --- reasonably well, again especially MVG.

The skewness pattern observed in the APOGEE data is not reproduced consistently well by either model.
We find a rise in skewness in the MVG velocities out to $l \sim 10^\circ$, at the same level as in the data,
but then the simulated high-velocity tails continue to much larger longitudes.
If these higher velocity stars are due to streaming motions along the bar (Section~\ref{sec:extragal}),
this could indicate a model bar that is longer than in the MW; although
the MVG bar has a total half-length of 4.5~kpc, which is consistent with the $4.6 \pm 0.3$~kpc thin bar of \citet[][]{Wegg_2015_RClongbar},
the orbital families giving rise to these structures are not tightly constrained and could vary significantly from the model
\citep[see also][]{Portail_2015_bulgeorbits}.

The skewness pattern in the S10 model is flatter, with a much smaller (perhaps insignificant) peak at $l \sim 10^\circ$.
The surface density of bar particles in Figure~\ref{fig:models}b reveals a fatter bar in the {\it XY} plane,
such that stars streaming along the far side at $l > 0^\circ$ are systematically farther away than in the MVG model.
If reasonably matched to the APOGEE sample, then, this may explain why the high-RV ``peaks'' are undetected 
in the S10 model \citep{Li_2014_nohighRVpeaks}.  
This model does reproduce better than MVG the ${\rm Skew}(V) \sim 0$ values in the inner disk, 
where the central thick bar ends and only the long bar remains ($l \gtrsim 20^\circ$).

Finally, both models match the flat ${\rm Kurt}(V) \approx 0$ values at all longitudes.

\subsection{Effect of Distance Limits} \label{sec:dist_cuts}

Contrary to Paper I, we do not apply distance restrictions to the plots shown above (beyond the removal of the dwarfs, necessarily closer to the Sun).
This choice was made because this work is intended as a step towards an accurate integrated-light 
``observation'' of the MW, which necessarily includes all of the stars.
We have repeated this analysis with the $4 < d < 12$~kpc limit imposed in Paper I,
and we find that the mean velocity, dispersion, and kurtosis patterns do not change significantly.

The largest change, when stars with $d<4$~kpc or $>$12~kpc are removed, is in the trend of skewness with longitude.
The high ${\rm Skew}(V)$ peak centered at $l \sim 10^\circ$ becomes much weaker, 
but we emphasize this is not because the closer disk stars are the ones at higher velocity.
In fact, the disk stars have the same velocity as the primary velocity peak in each field;
the more distant bar sample has a broader (but symmetric) distribution, 
overlapping the disk velocities but extending to higher velocities as well \citep[as described in, e.g.,][]{Aumer_2015_highRVbarstars}.
Removing the closer disk stars reduces the skewness by preferentially removing part of the lopsided distribution.
As discussed in Section~\ref{sec:high_velocity}, this may be the reason the so-called high-RV component is not
observed in data that are more confined to the innermost parts of the Galaxy.

\section{``High Velocity Component'' vs A Skewed Distribution} \label{sec:high_velocity}

Using APOGEE commissioning data taken towards the bulge, \citet{Nidever_2012_apogeebar} identified several lines of sight
whose velocity distributions showed a secondary high-velocity peak and could be decomposed into two overlapping gaussian distributions.
The primary peak followed the expected ``disk'' velocity, while the secondary peak had approximately constant velocity at all longitudes.
The Sgr dSph galaxy was ruled out as the source of the stars, and multiple kinematical models of the MW or MW-sized disk galaxies
could match the high-velocity components but not the trough that made them appear as peaks.

Subsequent efforts were made to identify these peaks in other datasets, 
including both observations and theoretical models scaled to the MW.  
Some reported detections \citep[e.g.,][]{Babusiaux_2014_barmetallicitykinematics},
some reported non-detections \citep[e.g.,][]{Li_2014_nohighRVpeaks,Zoccali_2014_GIBS1},
and others focused on explanations for the APOGEE data behavior 
\citep{Aumer_2015_highRVbarstars,Debattista_2015_kpcnucleardisk,Molloy_2015_highRVpeakorbits}.
For example, \citet{Debattista_2015_kpcnucleardisk} argue that the high velocity peaks 
are consistent with the presence of a rapidly rotating elliptical disk comprising stars on $x_2$ orbits, 
oriented perpendicular to the bar and extending $\sim$1~kpc in radius. 
In contrast, \citet{Molloy_2015_highRVpeakorbits} propose that the orbits of the stars in the high velocity peaks 
are dominated by those in the 2:1 $x_1$ orbit families that support the boxy/peanut morphology, 
which are most prominent in the plane due to the highest density of these orbits in this region.

In this section, we use the expanded DR12 dataset to demonstrate briefly that the initial detections are recoverable and verified,
and that this type of analysis can be applied to a larger sample of stars to obtain consistent results.
However, we show that the stars in this secondary ``component'' are not chemically distinct from the rest of the stars in the sample,
and thus we strongly argue that the identification of a secondary ``peak'' or distinct ``component'', and parameterization as such,
is a subjective and less robust way to characterize these LOSVDs.  
Use of the LOSVD skewness, as in Section~\ref{sec:high_moments} and Figure~\ref{fig:moments_plots}, 
is a more reliable way to quantify the details of the LOSVDs of all shapes, especially when comparing to models.

\begin{figure}[!hptb]
\begin{center}
  \includegraphics[trim=1.4in 1.1in 1.in 1.2in, clip, angle=180, width=0.45\textwidth]{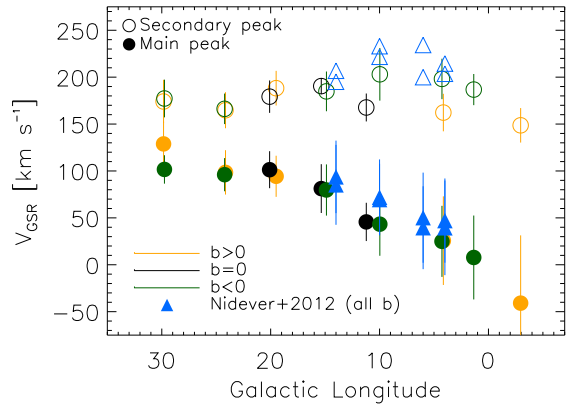} 
\end{center}
\caption{
Central velocities of the two velocity ``components'' in the APOGEE LOSVDs, as measured from a dual-gaussian fit. 
Filled circles indicate the primary peaks, open circles are the secondary peaks or tails, and triangles are the 
original measurements from \citet{Nidever_2012_apogeebar}.  
The new measurements from DR12 recover the original ones and are 
consistent with the trends predicted by those earlier detections. 
}
\label{fig:RV_peaks_velocity_vs_longitude}
\end{figure}

To demonstrate that the initial detections of \citet{Nidever_2012_apogeebar} are robust, 
we perform a comparable dual-gaussian fit to the expanded DR12 dataset, 
for stars with $|b| \le 5^\circ$ in bins with $\Delta l = 5^\circ$, $\Delta b = 2^\circ$.
The peak velocities of the best-fitting gaussians\footnote{The fits shown are those of bins with 
at least 50 stars and in which the LOSVD residuals of the dual-guassian fit have a mean $\le$4 and a dispersion $\le$4.  These criteria were found
by visual inspection to identify clear two-component fits.  Bins which did not pass these criteria could not be made to have a reasonable, convergent dual-gaussian representation.}
 are shown in Figure~\ref{fig:RV_peaks_velocity_vs_longitude} (circles), 
along with the derived velocities of the original ``components'' (triangles).
The results from the DR12 dataset are entirely consistent with the original detections, 
and data from the longitudes not available in 2012 lie along the same trends predicted by the earlier results.
Restricting the fitted sample to increasingly low $|b|$ results in more prominent (but noisier) high velocity tails,
suggesting that previous non-detections were impacted by the 
higher latitude coverage of those surveys (see also Figure~\ref{fig:compare_moments_plots}).

\begin{figure}[!hptb]
\begin{center}
  \includegraphics[trim=2.5in 1.6in 1.in 1.in, clip, angle=180, width=0.5\textwidth]{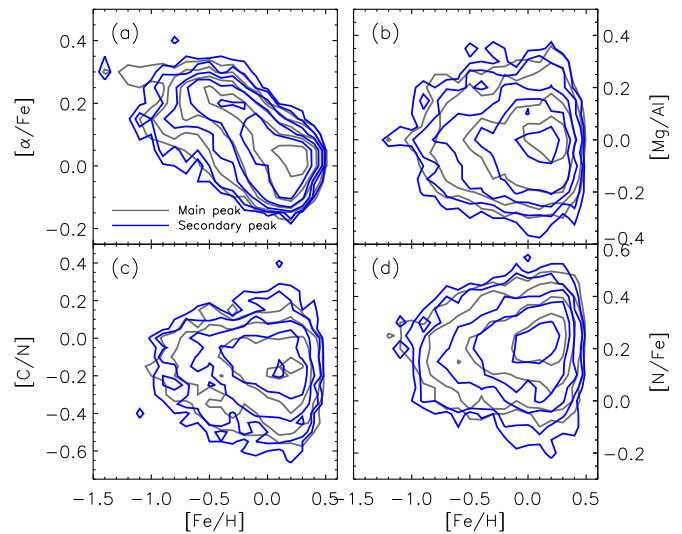} 
\end{center}
\caption{
Distribution of four sample abundances as a function of [Fe/H] for the ``main'' (blue) and ``high-RV'' (grey) peak stars: 
(a) [$\alpha$/Fe], (b) [Mg/Al], (c) [C/N], (d) [N/Fe].
No chemical differences are seen between these two kinematically-defined subsamples.
}
\label{fig:RV_peaks_velocity_vs_feh}
\end{figure}

In Figure~\ref{fig:RV_peaks_velocity_vs_feh}, we show
representative abundances, as a function of [Fe/H], for stars in the ``main'' peak (grey contours) and in the ``high-RV'' peak (blue contours).
For this figure, stars were assigned to the component they lie the fewest standard deviations away from, based on the dual gaussian fits.
The stars at higher velocities clearly have the same chemical patterns as those at lower velocities (see also Y.~Zhou et al., in prep).
{\it We conclude that the ``high-RV'' signature is real, but is not caused by a distinct stellar population
in the inner Galaxy; as such, using a metric like the skewness of the distribution is a more reliable way to describe the signature. }

\section{Cylindrical Rotation} \label{sec:cylindrical_rotation}

\begin{figure*}[]
\begin{center}
  \includegraphics[trim=1.3in 1.2in 0.8in 4.2in, clip, angle=180, width=0.95\textwidth]{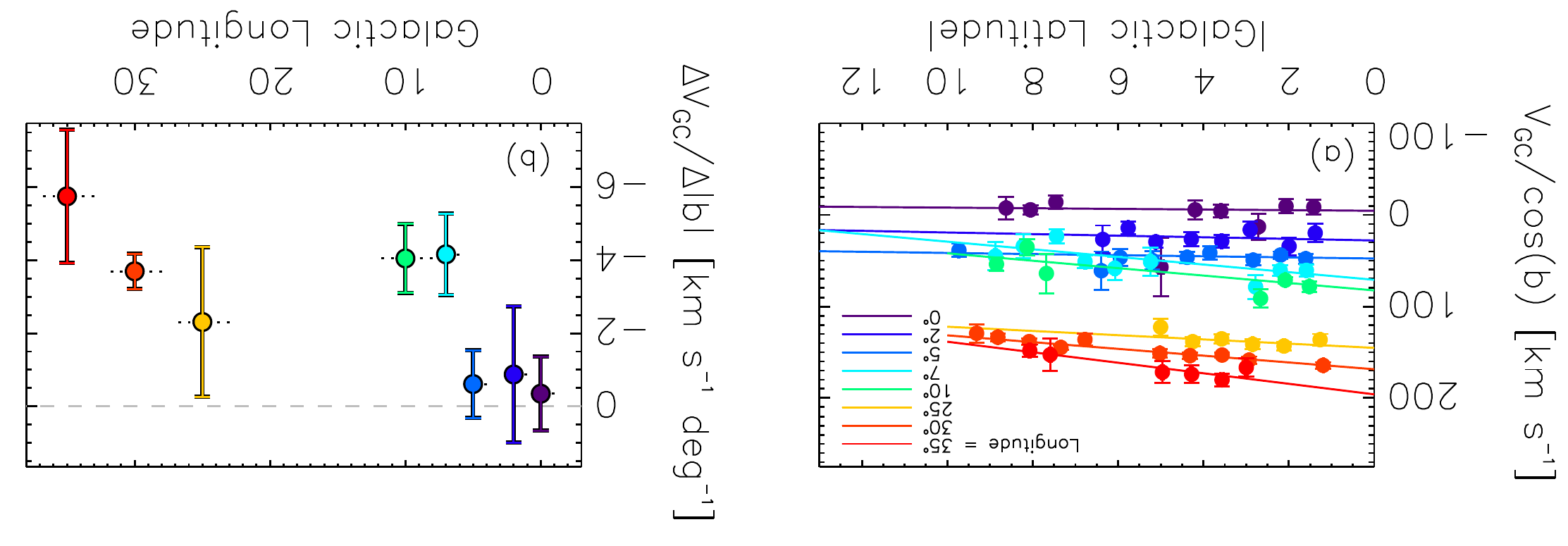} 
\end{center}
\vspace{-0.5cm}
\caption{
Measurement of cylindrical rotation, via dependence of mean velocity on latitude.
(a): $V_{\rm GC}$/$\cos{b}$ versus $|b|$ 
for different longitude bins.  
The data points are the mean velocities in each $\Delta |b| = 0.75^\circ$ latitude bin, and the error bars reflect the uncertainty in the mean.
(b): The slopes of the trends, as a function of longitude (same colors as in (a)).
The horizontal dotted error bars indicate the pre-set width of the longitude bins
($2^\circ < \Delta l < 4^\circ$).
Note the change in the vertical behavior of the rotation for $l \gtrsim 7^\circ$.
}
\label{fig:cylindrical_rotation}
\end{figure*}

Qualitatively, a nonspherical system that is {\it cylindrically rotating} has a mean rotation speed that is largely independent
of distance from the major axis. 
Cylindrical rotation is generally considered one of the hallmarks of a bar-dominated boxy/peanut bulge, 
both because it is a global kinematical pattern predicted by simulations of buckled bars 
\citep[e.g.,][]{Combes_1990_boxpeanutbars,Athanassoula_2002_nbodybars,Saha_2012_classicalbulgespinup,Shen_2016_theoreticalbulgemodels}, 
and because it has been observed in edge-on disk galaxies with boxy/peanut shaped central 
isophotes and other bar indicators \citep[e.g.,][]{Kormendy_1982_bulgerotation,FalconBarroso_2006_SAURONkinematics}.

{\it Quantitatively}, however, cylindrical rotation in external galaxies has been a more nebulous metric.
Part of the difficulty lies in confounding factors like dust or unknown bar angle, 
and part in the uncertain uniqueness and ubiquity of cylindrical rotation among boxy/peanut bars or bulges.
For example, \citet{Williams_2011_boxybulges} identified a range of rotation profiles in a small sample of boxy/peanut bulges,
\citet{Saha_2013_cylindricalrotation} demonstrated that classical bulges of merger origin could evolve into cylindrical rotation,
and \citet{Molaeinezhad_2016_cylindricalBProtation} observed cylindrical rotation in a few galaxies without visible boxy/peanut bulges.

A quantitative assessment of the ``cylindricality'' of the MW's rotation that can be compared to these external systems 
is a valuable datapoint.  Numerous surveys have qualitatively demonstrated that the mean velocity of the Milky Way's inner regions 
is very weakly dependent on latitude \citep[e.g.,][Paper I]{Howard_2009_bravarotation,Zoccali_2014_GIBS1}.
A numerical metric for cylindrical rotation was proposed by \citet{Saha_2013_cylindricalrotation} and 
refined (and applied to the MW) by \citet{Molaeinezhad_2016_cylindricalBProtation}.
Here, we perform a more finely grained assessment of the Milky Way's behavior, 
including extension into the disk beyond the known contribution of the central bar and 
using the more comprehensive angular coverage of the APOGEE dataset.

In Figure~\ref{fig:cylindrical_rotation}a, we show the mean $V_{\rm GC}$ (divided by $\cos{b}$ to account for projection effects)
as a function of latitude, for several longitude ranges, centered on the values indicated in the legend.
Also shown are the best-fitting straight lines for each longitude's points.
True cylindrical rotation in this representation would be reflected by a flat line ($\Delta V/ \Delta |b| = 0$).
The data were binned in latitude ($\Delta |b| = 0.75^\circ$) before fitting to prevent 
uneven latitude sampling from biasing the fits differently at different longitudes,
and the fitted latitude range ($1^\circ \le |b| \le 10^\circ$) was chosen after visual inspection of the stellar distance distributions
at different values of $|b|$, to ensure the velocity trend with latitude is not influenced by systematic distance shifts with latitude.

Figure~\ref{fig:cylindrical_rotation}b presents the slopes of these trends as a function of longitude (again, the dashed line
at $\Delta V/ \Delta |b| = 0$ indicates pure cylindrical rotation).
The longitude ranges vary in size, as the APOGEE spatial sampling is not completely uniform, and the width of each is indicated with the
horizontal uncertainty bars in Figure~\ref{fig:cylindrical_rotation}b.

For $l \lesssim 7^\circ$, the change in mean velocity with latitude is consistent with zero.
At higher longitudes, we see a vertical gradient, 
consistent with a smaller bulge contribution there, though we note that, interestingly, this begins within the 
generally accepted ``edge'' of the boxy bulge (at $l \sim 10^\circ$).
When fit within more restricted $|b|$ ranges ($|b| \leq 3^\circ, 4.5^\circ, 6^\circ$) to see if the larger longitudes are cylindrically rotating in the midplane,
we find noisier fits (unsurprisingly) but greater consistency with $\Delta V/ \Delta |b| = 0$ at most longitudes, especially $l < 10^\circ$.
This suggests the transition from cylindrical to non-cylindrical rotation occurs gradually, 
and preferentially at higher latitudes.

We calculated the cylindrical rotation metric $m_{\rm cyl}$\footnote{This method
entails fitting a straight line to the velocity-latitude trend, as we do here, but with velocities normalized to their midplane values.
The slope of these normalized fits are then averaged, $m_{\rm avg}$, and the final metric is  $m_{\rm cyl} = m_{\rm avg} + 1$.}
of \citet{Molaeinezhad_2016_cylindricalBProtation} using the same $|b|$ and $|l|$ ranges 
adopted for the GIBS sample in that paper (with the addition of our $|b| \ge 1^\circ$ limit described above), 
and found a value of $m_{\rm cyl} = 0.60 \pm 0.12$ for the APOGEE sample here.  
This is consistent with the \citet{Molaeinezhad_2016_cylindricalBProtation} value within 1$\sigma$, but we note that
this value is highly sensitive to the $l$ and $b$ limits and to the longitude binning for each linear fit.  
This metric also becomes significantly noisier at low $|l|$, where $V_{\rm GC}(b=0^\circ)$ (i.e., $V_{\rm max}$) is very small 
and thus the normalized velocity values $V/V_{\rm max}$ have a higher scatter.
For example, in a case where all longitudes had an identical tiny $\Delta V/ \Delta |b| < 0$, 
the bins at lower longitudes will yield a larger $m_{\rm cyl}$ than when the higher-longitude ones, with the same intrinsic behavior, are included.

We also assessed differences in the $\Delta V/ \Delta |b|$ behavior as a function of metallicity, 
using two bins defined with ${\rm [Fe/H}] \le -0.5$ and ${\rm [Fe/H]} \ge -0.25$, to highlight any differences as much as possible.
Repeating the analysis described above for each bin, we observed that the metal-rich subsample had nearly identical
behavior as the total sample (as expected, given that most of the stars are metal-rich).

The fits for the metal-poor subsample are significantly noisier due to the reduced numbers of stars, 
and not able to be measured reliably at all longitudes,
but the slopes are still generally consistent with zero within $l \lesssim 7^\circ$ 
and negative at levels similar to the total sample outside of $l \sim 7^\circ$.  
The exception is the stack of fields at $l=30^\circ$, which contains relatively few stars in the metal-poor bin, 
but those show a clearly flat slope in $\Delta V/ \Delta |b|$ (where the metal-rich and total samples have the steepest slope).
This is certainly not a contribution from the central bar here, nor from the long bar at these latitudes.
This may be a signature of a large thick disk contribution,
but in any event, it serves as an example where ``cylindrical rotation'' $\neq$ ``bar''.

\section{Extragalactic Bar/Bulge Diagnostics} \label{sec:extragal}

In a massive edge-on disk galaxy, the critical size and shape characteristics of the central bar/bulge are often difficult to define photometrically.
Sometimes the light distribution just outside the plane can be used to identify a boxy or peanut bulge, suggesting a bar, 
but disk extinction confounds measurement of the midplane regions and any thin planar structures \citep[e.g.,][]{Bureau_1999_bardiagnostics}.

Kinematical signatures, with or without high-quality photometry, can provide a clearer view of the signatures
of a bar by measuring its impact on stars throughout the inner galaxy \citep[e.g.,][]{Kuijken_95_peanutbars}.
These kinematical diagnostics include comparison to position-velocity diagrams \citep[PVDs;][]{Kuijken_95_peanutbars} and velocity moments
informed by hydrodynamical simulations \citep{Athanassoula_1999_hydrobardiagnostics} or 
$N$-body simulations \citep{Bureau_05_nbodybardiagnostics,Iannuzzi_2015_2D_BPbulges}.

\begin{figure}[!hptb]
\begin{center}
  \includegraphics[trim=6in 1.5in 1in 1.5in, angle=180, width=0.45\textwidth]{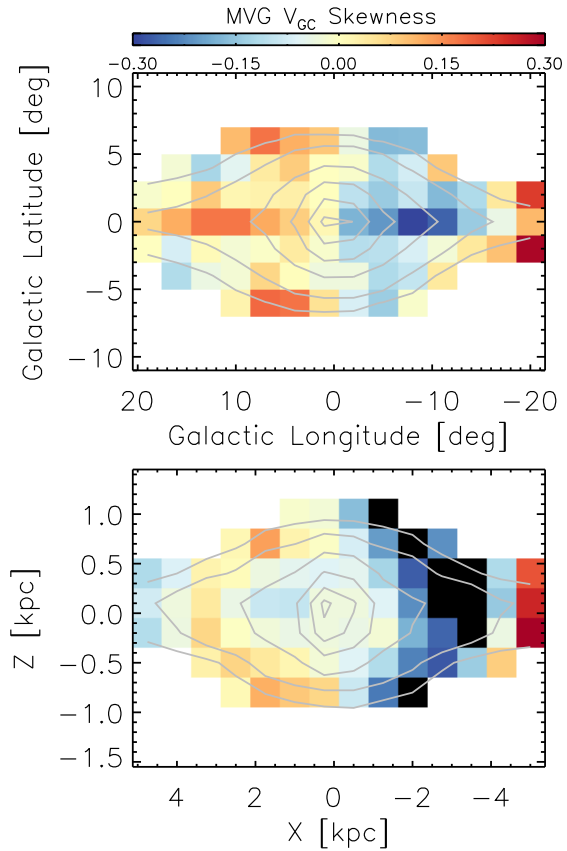} 
\end{center}
\caption{
Maps of the velocity skewness
extracted from a MW-like $N$-body simulation \citep[MVG;][]{MartinezValpuesta_2013_barmodel},
in two projections: $(l,b)$ and $(X,Z)$.
Gray contours indicate the projected particle densities.
This comparison highlights one difference between studying galaxies from internal and external perspectives.
See text for details.
}
\label{fig:model_perspective}
\end{figure}

The MW is the best place to test these diagnostics, where we can understand whether the assumed stars and stellar orbits
giving rise to the patterns seen in integrated light are borne out by observations of the resolved stars themselves.
The comparison between MW data and observations or simulations of external galaxies is not immediately straightforward.
In addition to the caveats described in Section~\ref{sec:low_moments}, 
the kinematical (and chemical, etc.) patterns observed as a function of $(l,b)$ within the MW may differ from what
would be observed as a function of Galactocentric $X$, $Y$, or $Z$.
Figure~\ref{fig:model_perspective} shows an example of this, in which ${\rm Skew}(V)$ is computed for the MVG simulation particles 
(Section~\ref{sec:model_data} and Figure~\ref{fig:models})
inside the solar circle as a function of $(l,b)$ and of $(X,Z)$.  The former mimics our perspective of the MW, 
and the latter is similar to our perspective of other disk galaxies, with a bar at the same angle.
Notice that both projections display skewness patterns driven by the bar, but these are slightly different: in $(l,b)$, the strongest
signature is in the midplane, with a broad, positive skewness peak centered on the end of the near side of the bar 
(as seen in, e.g., Figure~\ref{fig:moments_plots}) and a corresponding negative skewness peak towards the bar's far end.
In the $(X,Z)$ projection, positive and negative skewness is induced {\it off} the plane in the innermost kpc
\citep[see also][]{Iannuzzi_2015_2D_BPbulges}, 
with the strongest midplane signatures restricted to narrow skewness peaks at the ends of the bar.

\begin{figure*}[!hptb]
\begin{center}
  \includegraphics[trim=0.6in 1.5in 0.8in 4.5in, clip, angle=180, width=0.95\textwidth]{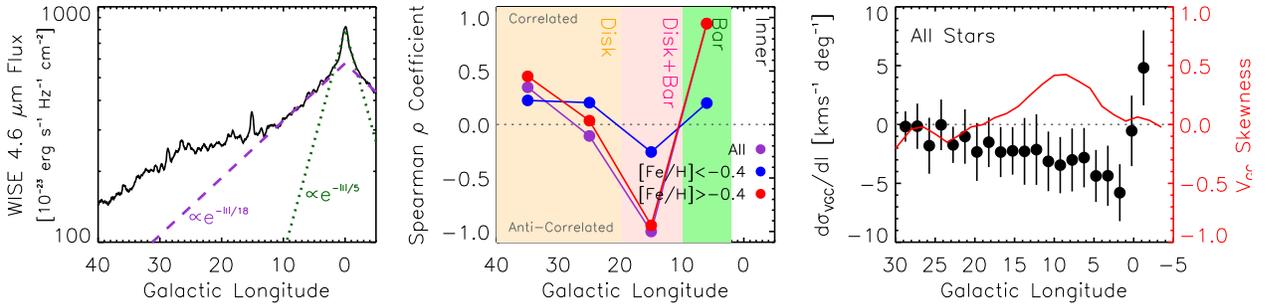} 
\end{center}
\vspace{-0.5cm}
\caption{
Evaluation of WISE flux and APOGEE kinematical properties in light of end-on bar diagnostics supported by {\it N}-body simulations
and observations of external galaxies (see details in Section~\ref{sec:extragal}).
{\it Left:} The [4.6$\mu$] flux from WISE (summed within $|b| \le 5^\circ$), fitted with two exponential profiles.
{\it Center:} The correlation between mean velocity and velocity skewness for different regimes of the inner MW, 
for the metal-rich and metal-poor bins used in this analysis.
{\it Right:} The relationship between the velocity skewness and the change in the velocity dispersion with longitude.
}
\label{fig:compare_diagnostics}
\end{figure*}

Provided we bear these considerations in mind,
we can use some of the qualitative features of the kinematical patterns observed in the APOGEE sample to begin 
to confront the predictions of barred simulations with observations of the barred MW.
\citet{Bureau_05_nbodybardiagnostics} provide some of these feature predictions for strong bars seen roughly end-on,
based on $N$-body simulations of barred disks (e.g., their Figure~1).  
These qualitative diagnostics are reproduced in the numbered, italicized text below,
and we show some of the comparisons with APOGEE data in Figure~\ref{fig:compare_diagnostics}.
 
(1) {\it Light profile along the major axis with a quasi-exponential central peak and a plateau at moderate radius}, where ``moderate radius'' refers to the end(s) of the primary central bar.  In the left panel of Figure~\ref{fig:compare_diagnostics}, we show the WISE 4.6$\mu$m flux density \citep[$|b| \leq 5^\circ$;][]{Lang_2014_unWISEcoadds} fitted with two exponential trends within the range of the bar: $|l| \leq 2^\circ$ and $3^\circ \leq |l| \leq 10^\circ$.  The very innermost range --- the center of the MW --- has a density profile discontinuous with the bulk of the bulge, which has been attributed to distinct morphological features \citep[e.g.,][but see also \citet{Gerhard_2012_nuclearbar}]{Alard_01_centralbar,Robin_2012_besanconbarmodel}.  Due to the high extinction and small angular span of this regime, the stars responsible for this flux feature are unlikely to be well-represented in our spectroscopic sample.  Focusing on the truly bar-dominated range ($3^\circ \leq |l| \leq 10^\circ$), where the observed kinematical patterns are actually connected to the observed integrated flux, we indeed see an exponential flux density decrease with longitude out to $l \sim \pm10^\circ$ (only the first quadrant is shown here).  This is the canonical end of the central bar, and beyond this location the flux has a more gradual plateau to past $l \sim 30^\circ$.

(2) {\it A double-peaked rotation curve,} where the first local ``minimum'' occurs at the base of the central flux peak.  Bearing in mind the caveat above about the disconnect between WISE flux and APOGEE kinematics within $|l| \lesssim 3^\circ$, we point to the $<$$V_{\rm GC}$$>$$(l)$ panel of Figure~\ref{fig:compare_moments_plots} (which shows the mean velocity sampled more finely in $l$ than Figure~\ref{fig:moments_plots}).  Based on point (1) above, the base of the central flux peak may be identified around $|l| \sim 2-3^\circ$, where the two exponential profiles have comparable contributions.  At this point in Figure~\ref{fig:compare_moments_plots}, the $<$$V_{\rm GC}$$>$ of the APOGEE sample (represented by the dominant metal-rich stars, shown in pink) does appear to flatten, before increasing again from $l \sim 5^\circ$ outwards.

(3) {\it A central velocity dispersion peak with a plateau at moderate radius; occasionally the very center is flat-topped or even shows a minimum, and there may be secondary peaks outside of the center.}  In Figures~\ref{fig:moments_plots} and \ref{fig:compare_moments_plots}, we see a central, singly peaked dispersion that decreases monotonically, with a flattening slope, into the disk.  This flattening occurs more rapidly in the metal-poor stars, though both populations have parallel plateaus by $l \sim 30^\circ$.

(4) {\it A correlation between the skewness and the mean velocity over the projected bar length.}  
\citet{Iannuzzi_2015_2D_BPbulges} note that the correlation persists until ``the end of the [Boxy/Peanut] structure'', and \citet{Bureau_05_nbodybardiagnostics} also describe a skewness-velocity {\it anti}-correlation beyond the end of the bar, before returning to zero or positive correlation.  In the middle panel of Figure~\ref{fig:compare_diagnostics}, we show the Spearman correlation coefficient $\rho$ between the mean velocity and the skewness in a number of longitude ranges dominated by different components.  In the $3^\circ \leq l \leq 10^\circ$ range dominated by the central bar (green shading), we see a positive correlation between skewness and mean velocity, in the metal-richer stars only (red points, as in Section~\ref{sec:high_moments}).  At and beyond the end of the bar dominance (say, $l = 10^\circ-20^\circ$, pink shading), we indeed see a skew-velocity anti-correlation, again only in the more metal rich stars, before zero and small (positive) correlations in the regions dominated by the inner disk (orange shading).  

\citet{Iannuzzi_2015_2D_BPbulges} note that the skewness peak coincides with a plateau in the velocity dispersion.  In the right panel of Figure~\ref{fig:compare_diagnostics}, we show both the slope of the velocity dispersion ($d\sigma_{\rm V} / dl$, black points) and the skewness (red line) as a function of longitude.  The change in dispersion with longitude decreases at larger longitudes (i.e., the dispersion becomes increasingly plateau-like), with the most dramatic slope changes at small longitudes near the rise of the skewness peak ($l \sim 5^\circ$).  Thus the skewness peak itself, around $l \sim 10^\circ$, does coincide with the transition to a flatter dispersion, though the feature is not dramatic.  We remind the reader that this skewness peak appears only in the metal-richer subsample, so the statement that the kinematical behavior is qualitatively similar to that of the barred galaxies of \citet{Iannuzzi_2015_2D_BPbulges} does not hold true for the MW bulge's metal-poorer subsample.

Finally, we note that there are a number of extragalactic bar signatures described in these works that are not yet testable with APOGEE. 
For example, \citet[][]{Iannuzzi_2015_2D_BPbulges} describe the skew and kurtosis features that appear in velocity distributions {\it off} the midplane
in the presence of a boxy/peanut bulge (see also Figure~\ref{fig:model_perspective} above).  
We do not yet have a dense sampling of stars over a large enough contiguous span of these off-midplane regions to measure this.
As data accumulate from APOGEE-2 and other inner Galaxy surveys, 
these diagnostics measured in the MW will provide very interesting tests of both barred galaxy models 
and our interpretations of integrated light velocity fields in external systems.

\section{Summary} \label{sec:finis}
Understanding the finely detailed behavior of stars in the inner MW, including the relationships between their chemical and kinematical patterns, 
is critical not only to retracing the evolutionary history of the MW, but also to interpreting the unresolved inner regions of external galaxies
and placing the MW in context with those systems.
Along with the line-of-sight velocity mean and dispersion, 
we present here the first maps of the velocity skewness and kurtosis of inner Galaxy stars, 
independently for relatively metal rich (${\rm [Fe/H]} > -0.4$) and metal poor (${\rm [Fe/H]} \ge -0.4$) stars.
We find patterns that, based on comparison to $N$-body simulations,  
are consistent overall with the metal rich stars being more representative of stars entrained in the bar --- 
e.g., an increase in the velocity skewness at the projected end of the bar --- 
and the metal poorer stars being kinematically hotter and less bar-like.

We revisit the issue of the high velocity ``peaks'' in the inner Galaxy and verify the initial detections \citep{Nidever_2012_apogeebar}, 
but we conclude that the velocity skewness is a significantly more robust way 
to quantify this phenomenon than the fitting of discrete components to the velocity distribution.
This conclusion is supported by the chemical similarity of the ``high velocity'' stars to the rest of the sample,
which indicates these are not distinct and separable populations.
We explore the extent of cylindrical rotation in our sample and find an apparent break in the pattern between $l=7^\circ$ and $l=10^\circ$, 
near the end of the boxy bulge, as expected.  

Throughout this work, we also discuss the advantages, disadvantages, and caveats 
of using MW data to interpret unresolved stellar behavior in external galaxies.
Some qualitative diagnostics proposed to identify end-on bars in these galaxies are compared to the patterns we see in the MW,
which we know from many lines of evidence to show a nearly end-on bar towards the Sun.
These diagnostics include the midplane integrated flux profile and the correlation between mean velocity and velocity skewness.
We find good agreement in these qualitative comparisons, and 
in future work, we will explore these questions quantitatively, framing the MW as
an unparalleled Rosetta Stone for the signatures of chemodynamical evolution 
observed in galaxies throughout the Universe.

\begin{acknowledgments}

GZ has been supported by an NSF Astronomy \& Astrophysics Postdoctoral Fellowship under Award No.\ AST-1203017,
and thanks the Max-Planck-Institut f\"ur Astronomie for hospitality that advanced the progress of this paper.
JAJ acknowledges support from NSF AST-1211853.
We also thank Z.-Y.~Li, J.~Shen, and D.L.~Nidever for very helpful discussions,
and R.~P.~Schiavon and the anonymous referee for suggestions that improved the clarity of the manuscript.

Funding for SDSS-III has been provided by the Alfred P. Sloan Foundation, the Participating Institutions, the National Science Foundation, and the U.S. Department of Energy Office of Science. The SDSS-III web site is \url{http://www.sdss3.org/}.

SDSS-III is managed by the Astrophysical Research Consortium for the Participating Institutions of the SDSS-III Collaboration including the University of Arizona, the Brazilian Participation Group, Brookhaven National Laboratory, University of Cambridge, Carnegie Mellon University, University of Florida, the French Participation Group, the German Participation Group, Harvard University, the Instituto de Astrof\'{i}sica de Canarias, the Michigan State/Notre Dame/JINA Participation Group, Johns Hopkins University, Lawrence Berkeley National Laboratory, Max Planck Institute for Astrophysics, Max Planck Institute for Extraterrestrial Physics, New Mexico State University, New York University, The Ohio State University, Pennsylvania State University, University of Portsmouth, Princeton University, the Spanish Participation Group, University of Tokyo, University of Utah, Vanderbilt University, University of Virginia, University of Washington, and Yale University. 

\end{acknowledgments}

\bibliographystyle{aasjournal}
\bibliography{/Users/GailZasowski/Documents/APOGEE/analysis/reflib}

\end{document}